\documentclass[12pt]{iopart}
\usepackage{iopams}                            
\usepackage{subfigure}
\usepackage{graphicx}
\usepackage{color}
\usepackage{subeqn}                            
\usepackage{cite}
\usepackage[en-US,showzone=false]{datetime2}       
\newcommand{\Schrodinger}{Schr{\"o}dinger}     
\newcommand{\ansatz}{ans{\"a}tz}               
\newcommand{\PT}{\mathcal{PT}}                 
\newcommand{\SUSY}{\mathcal{SUSY}}             
\newcommand{\eqref}[1]{(\ref{#1})}             
\newcommand{\ef}[1]{(\ref{#1})}                
\eqnobysec                                     
\newcommand{\dd}[1]{\,\mathrm{d}#1\,}
\newcommand{\ddd}[1]{\,\mathrm{d}^{2}#1\,}
\newcommand{\sech}{\,\mathrm{sech}}            
\newcommand{\Imag}[1]{ \,\mathrm{Im}{\left \{ \,#1\,\right \}}}   
\newcommand{\tint}{\!\int\!}                   
\newcommand{\tpsi}{\tilde{\psi}}               
\newcommand{\dv}[2]{\frac{\mathrm{d}#1}{\mathrm{d} #2}}
\newcommand{\pdv}[2]{\frac{\partial #1}{\partial #2}}
\newcommand{\ppdv}[2]{\frac{\partial^2 #1}{\partial {#2}^2}}
\newcommand{\qc}{\>,\quad}

\newcommand{\notag}{\nonumber}                  
\begin{document}
%
%
%
\title[NNLSE]{Stability of exact solutions of a nonlocal and nonlinear \Schrodinger\ equation with arbitrary nonlinearity}
\author{Efstathios G. Charalampidis$^1$, Fred Cooper$^{2,3}$, Avinash Khare$^4$, John F. Dawson$^5$, Avadh Saxena$^3$}
\address{$^1$Mathematics Department, California Polytechnic State University, San Luis Obispo,
CA 93407-0403, United States of America}   
\address{$^2$The Santa Fe Institute, 1399 Hyde Park Road, Santa Fe, NM 87501, United States of America}
\address{$^3$Theoretical Division and Center for Nonlinear Studies, Los Alamos National Laboratory, 
Los Alamos, NM 87545, United States of America}
\address{$^4$Physics Department, Savitribai Phule Pune University, Pune 411007, India}
\address{$^5$Department of Physics, University of New Hampshire, Durham, NH 03824, 
United States of America}
\ead{echarala@calpoly.edu}
\ead{cooper@santafe.edu}
\ead{avinashkhare45@gmail.com}
\ead{john.dawson@unh.edu}
\ead{avadh@lanl.gov}
\begin{abstract}
This work focuses on the study of solitary wave solutions to a nonlocal, 
nonlinear \Schrodinger\ system in $1$+$1$ dimensions with arbitrary 
nonlinearity parameter $\kappa$. Although the system we study here was
first reported by Yang (Phys.~Rev.~E, 98 (2018), 042202) for the fully 
integrable case $\kappa=1$, we extend its considerations and offer criteria 
for soliton stability and instability as a function of $\kappa$. In particular,
we show that for $\kappa <2$ the solutions are stable whereas for $\kappa >2$ 
they are subject to collapse or blowup. At the critical point of $\kappa=2$, there 
is a critical mass necessary for blowup or collapse. Furthermore, we show there 
is a simple one-component nonlocal Lagrangian governing the dynamics of the 
system which is amenable to a collective coordinate approximation. To that end, 
we introduce a trial wave function with two collective coordinates to study the 
small oscillations around the exact solution. We obtain analytical expressions 
for the small oscillation frequency for the width parameter in the collective 
coordinate approximation. We also discuss a four collective coordinate approximation 
which in turn breaks the symmetry of the exact solution by allowing for translational 
motion. The ensuing oscillations found in the latter case capture the response of 
the soliton to a small translation. Finally, our results are compared with numerical
simulations of the system.
\end{abstract}
\vspace{10pt}
\begin{indented}
\item[]{\DTMnow\ EST}
\vspace{5pt}
\item[] {LA-UR-21-20590}
\end{indented}
%
%
\submitto{\jpa}
\vspace{2pc}
\section{\label{s:Intro}Introduction}

The nonlinear \Schrodinger\ equation (NLSE) arises in many areas of 
physics including Bose-Einstein condensation, plasmas, water waves 
and nonlinear optics~\cite{Kevrekidis:2016aa}, among many others. The 
possibility of experimentally coupling two-component NLSEs in matrix 
complex potentials has recently been investigated in nonlinear optics 
situations in which two wave guides are locally coupled through an antisymmetric 
medium~\cite{PhysRevA.99.013823}. In~\cite{Cooper:2017aa,1751-8121-50-48-485205}, 
we studied the stability of exact solutions of a single component NLSE in 
a class of external potentials having supersymmetry ($\SUSY$) and parity-time, i.e., $\PT$ symmetry. 
We then extended our results to two-component NLSEs in $\PT$-symmetric 
and supersymmetric external potentials in Refs.~\cite{Charalampidis_2020,Charalampidis_2021}.  
The two-component system we studied in~\cite{Charalampidis_2020} in the absence 
of the external potential was a particular example of the Manakov system~\cite{Manakov:1973aa} 
now being studied in the context of nonlocal nonlinear \Schrodinger\ equations (NNLSEs).
The NNLSE, its variants and soliton solutions, have been studied in a variety of contexts~\cite{Ablowitz:2013aa,Ablowitz_2016,N1,STALIN2020126201,N3,N4,N5,PhysRevE.98.042202,N7,N8,N9,N10,N11,N12,N13,N14,N15}.

Here, and upon following Yang's proposal of imposing the solution constraint $\psi_2(x,t) = \psi_1(-x,t)$,
thus rendering the system to become \textit{nonlocal} (see~\cite{PhysRevE.98.042202}), we generalize these 
considerations by introducing an arbitrary nonlinearity with exponent denoted as $\kappa$ hereafter. 
It should be noted in passing that the resulting system is integrable only for $\kappa=1$~\cite{PhysRevE.98.042202}. 
In particular, we extend our previous discussion of two coupled NLSEs to the present case in order 
to compare the nonlocal stability results with those known for the usual one- and two-component 
\textit{local} NLSEs. A major difference in the solution space is that when we impose the above mentioned 
constraint, there are no longer moving single soliton solutions; instead they are trapped at the origin. 
To study the effect of small distortions of the initial solution, we use a variational approximation as 
well as perform numerical simulations. Small perturbations on an exact solution cause a slight increase 
in the energy. We find that the domains of stability in terms of the parameter $\kappa$ are the same as 
those found for the solitons in the usual Manakov system, where instability occurs for $\kappa > 2$. The 
collective coordinate (CC) approach gives qualitative agreement for the motion of the perturbed soliton with 
what is found in numerical simulations of the NNLSE. 

The structure of the paper is as follows. In Sec.~\ref{s:SchEq} we present our generalized model and give 
the exact yet trapped one-soliton solution to the coupled equations. In Sec.~\ref{s:action}, we discuss 
the derivation of the equations of motion from an action principle. The exact solution is given in Sec.~\ref{s:exact}, 
and conservation laws resulting from the action are presented in Sec.~\ref{s:conserved}. In Sec.~\ref{s:Derrick} 
we use both Derrick's theorem and the Vakhitov-Kolokokov (V-K) stability criterion to show that for $\kappa > 2$ 
the solutions are unstable. In Sec.~\ref{s:2CC} we introduce a 2CC variational approximation and give the 
equations of motion for these CCs. In Sec.~\ref{s:LinerResponse} we derive the linear response 
approximation to the CC equations and obtain the small oscillation frequency for the width 
parameter. The analysis of these equations shows that the soliton is unstable to small oscillations of the width 
when $\kappa >2$, a finding that is in full agreement with Derrick's theorem and the V-K criterion. In Sec.~\ref{s:blowup}, 
we give our results for the time evolution in the unstable regime as well as the critical mass when $\kappa=2$ in 
the CC approximation. In Sec.~\ref{s:4CC}, we consider a 4CC variational approximation and 
derive the associated equations of motion. The initial coordinate therein is displaced from the exact solution, 
and we compare the variational results with numerical simulations of the NNLSE. In Sec.~\ref{s:NumSim} we explain 
our numerical approach and then compare the numerical results to those of the CC approximation. 
Finally we state our conclusions in Sec.~\ref{s:conclusions}. 
%
%
%
\section{\label{s:SchEq}Yang's version of the nonlocal and nonlinear \Schrodinger\ equation}

In~\cite{PhysRevE.98.042202}, special solutions ($\kappa=1$) of the (generalized) 
Manakov system:
\begin{subeqnarray}\label{Manakov}
   \bigl \{ \,
      \rmi \, \partial_t 
      +
      \partial_x^2
      + 
      2 \, g \, [\, |\psi_1(x,t)|^2 + |\psi_2(x,t)|^2 \, ]^{\kappa} \,
   \bigr \} \, \psi_1(x,t)
   &=
   0 \>,
   \label{Manakov1} \\
   \bigl \{ \,
      \rmi \, \partial_t 
      +
      \partial_x^2
      + 
      2 \, g \, [\, |\psi_1(x,t)|^2 + |\psi_2(x,t)|^2 \, ]^{\kappa} \,
   \bigr \} \, \psi_2(x,t)
   &=
   0 \>
   \label{Manakov2} 
\end{subeqnarray}
were studied. Upon imposing the solution constraint
\begin{equation}\label{constraint}
   \psi_2(x,t) = \psi_1(-x,t) \>,
\end{equation}
Eqs.~\eqref{Manakov1}-\eqref{Manakov2} reduce to the single 
nonlinear and nonlocal \Schrodinger\ equation (NNLSE) of the form:
\begin{equation}\label{NNLSE}
   \bigl \{\,
      i \,
      \partial_t
      +  
      \partial_x^2
      +
      2 \, g \, [\, |\psi(x,t)|^2 + |\psi(-x,t)|^2 \, ]^{\kappa}
   \bigr \} \, \psi(x,t) 
   = 
   0 \>,
\end{equation}
whose Lax pair for $\kappa = 1$~\cite{Yang_book} reads
\begin{subeqnarray}\label{LaxPair}
   \fl
   \partial_x \, Y(x,t)
   &=
   \bigl \{\, - \rmi \, \zeta \, J + Q(x,t) \, \bigr \} \, Y(x,t) \>,
   \label{LaxPair1} \\
   \fl
   \partial_t Y(x,t)
   &=
   \bigl \{\, 
      - 
      2 \rmi \, \zeta^2 \, J 
      + 
      2 \, \zeta \, Q(x,t)
      +
      \rmi \, J \, [\, \partial_x Q(x,t) - Q^2(x,t) \,] \,
   \bigr \} \, Y(x,t) \>
\end{subeqnarray}
together with
\begin{equation}\label{JQdefs}
   \fl
   J
   =
   \left (
   \begin{array}{ccc}
      1 & 0 & 0 \\
      0 & 1 & 0 \\
      0 & 0 & -1
   \end{array}
   \right ) \>,
   \qquad
   Q(x,t)
   =
   \left (
   \begin{array}{ccc}
      0 & 0 & \psi_1(x,t) \\
      0 & 0 & \psi_2(x,t) \\
      - g \, \psi_1^{\ast}(x,t) & - g \, \psi_2^{\ast}(x,t) & 0
   \end{array}
   \right ) \>.   
\end{equation}
We note in passing that $Y(x,t)$ is the Lax vector in the zero curvature representation 
and $\zeta$ is the spectral function. The Lax pair for Eq.~\eqref{NNLSE} is obtained by 
applying Eq.~\ef{constraint} to Eq.~\ef{JQdefs}, and its existence renders the system to be
integrable for $\kappa = 1$ (even though this is a Hamiltonian system for all $\kappa$, 
it is integrable only for $\kappa = 1$).

Using this method, Yang in~\cite{PhysRevE.98.042202} found not only one-soliton solutions
of the form:
\begin{equation}\label{YangOneSoliton}
   \psi(x,t)
   =
   \frac{\beta}{\sqrt{2}} \, \sech(\beta x) \,\rme^{\rmi \, \beta^2 \, t} \>
\end{equation}
but also two- and three- soliton solutions of the NNLSE, all for $g = 1$ (the focusing case) 
and $\kappa = 1$ (integrable case). It should be noted that the constraint given by 
Eq.~\eqref{constraint} is different from the one first suggested by Ablowitz and Musslimani 
in~\cite{Ablowitz:2013aa,Ablowitz_2016}; they considered a different system consisting of two 
coupled nonlinear \Schrodinger\ equations (NLSEs), namely:
\begin{subeqnarray}\label{Ablowitz}
   \bigl \{ \,
      \phantom{-}
      \rmi \, \partial_t 
      +
      \partial_x^2
      + 
      2 \, g \, \psi_1(x,t) \, \psi_2(x,t) \,
   \bigr \} \, \psi_1(x,t)
   &=
   0 \>,
   \label{Ablowitz1} \\
   \bigl \{ \,
      -
      \rmi \, \partial_t 
      +
      \partial_x^2
      + 
      2 \, g \, \psi_2(x,t) \, \psi_1(x,t) \,
   \bigr \} \, \psi_2(x,t)
   &=
   0 \>.
   \label{Ablowitz2} 
\end{subeqnarray}
Imposing the solution constraint: $\psi_2(x,t) = \psi_1^\ast(-x,t)$, Eqs.~\eqref{Ablowitz1}-\eqref{Ablowitz2}
reduce to the single nonlocal and nonlinear equation:
\begin{equation}\label{AblowitzII}
   \bigl \{ \,
      \rmi \, \partial_t 
      +
      \partial_x^2
      + 
      2 \, g \, \psi(x,t) \, \psi^{\ast}(-x,t) \,
   \bigr \} \, \psi(x,t)
   =
   0 \>.
\end{equation}
The advantage of the system proposed by Yang is that it is accessible in nonlinear optics,
and the first three conserved quantities are real by construction. Also the  stability of 
the solutions can be studied using techniques we used to study the usual two-component NLSEs~\cite{Cooper:2017aa}.
%
%
\section{\label{s:action}Action principle} 

The Manakov system of Eq.~\eqref{Manakov} can be written in vector form 
as
\begin{equation}\label{Manakov2D}
   \bigl \{ \,
      \rmi \, \partial_t 
      +
      \partial_x^2
      + 
      2 \, g \, [\, \Psi^{\dag}(x,t) \, \Psi(x,t) \,]^{\kappa} \,
   \bigr \} \, \Psi(x,t)
   =
   0 \>,
\end{equation}
with
\begin{equation}\label{Psidef}
   \Psi(x,t)
   =
   \Bigl ( \begin{array}{c}
      \psi_1(x,t) \\
      \psi_2(x,t)
   \end{array} \Bigr )\in\mathbb{C}^{2}\>.
\end{equation}
It can be shown that Eq.~\eqref{Manakov2D} can be derived from 
an action principle. Indeed, let
\begin{equation}\label{Action2D}
   \Gamma[\Psi^{\dag},\Psi]
   =
   \int \dd{t} L[\Psi^{\dag},\Psi] \>
\end{equation}
be the action of the system where $L$ stands for its 
Lagrangian given by
\begin{equation}\label{Lagrangian}
   L[\Psi^{\dag},\Psi]
   =
   T[\Psi^{\dag},\Psi] - H[\Psi^{\dag},\Psi] \>,
\end{equation}
with
\begin{subeqnarray}\label{TandH}
   T[\,\Psi^{\dag},\Psi\,]
   &=
   \tint \dd{x} 
   \frac{\rmi}{2} \,
   \biggl \{\,
      \Psi^{\dag}(x,t) [\, \partial_t \Psi(x,t) \,]
      -
      [\, \partial_t \Psi^{\dag}(x,t) \,] \, \Psi(x,t) \,
   \biggr \} \>,
   \label{Tdef} \\
   H[\,\Psi^{\dag},\Psi\,]
   &=
   \tint \dd{x} 
   \Bigl \{\,
      | \, \partial_x \Psi(x,t) \, |^2
      -
      \frac{2 \, g}{\kappa+1} \, [\, \Psi^{\dag}(x,t) \Psi(x,t) \,]^{\kappa+1}
   \Bigr \} \>.   
   \label{Hdef}   
\end{subeqnarray} 
Once we impose the constraint of Yang, then Eq.~\eqref{NNLSE} can be obtained 
from the following nonlocal yet one-component action principle:
\begin{eqnarray}\label{ActionDefs}
   S[\psi,\psi^{\ast}]
   &=
   \tint \dd{t} L[\psi,\psi^{\ast}] 
   \qc \{\, \psi,\psi^{\ast} \,\} \in \mathbb{C} \>,
   \label{Action} \\
   L[\psi,\psi^{\ast}]
   &=
   T[\psi,\psi^{\ast}] - H[\psi,\psi^{\ast}] \>,
   \label{LagrangianII}
\end{eqnarray}
with
\begin{subeqnarray}\label{TandHt}
   \fl
   T[\,\psi,\psi^{\ast}\,]
   &=
   \frac{\rmi}{2}
   \tint \dd{x} 
   \biggl \{\,
      \psi^{\ast}(x,t) [\, \partial_t \psi(x,t) \,]
      -
      [\, \partial_t \psi^{\ast}(x,t) \,] \, \psi(x,t) \,
   \biggr \} \>,
   \label{Tvalue} \\
   \fl
   H[\,\psi,\psi^{\ast}\,]
   &=
   \tint \dd{x} 
   \Bigl \{\,
      | \, \partial_x \psi(x,t) \, |^2
      -
      \frac{g}{\kappa+1} \, 
      [\, 
         |\psi(x,t)|^2 + |\psi(-x,t)|^2 \,
      ]^{\kappa+1}
   \Bigr \} \>.   
   \label{Hvalue}   
\end{subeqnarray}
This way, the Lagrange's equation of motion
\begin{equation}\label{LagEqu}
   \frac{\delta L[\psi,\psi^{\ast}]}{\delta \psi^{\ast}(x,t)}
   -
   \dv{}{t} 
   \Bigl [\, 
      \frac{L[\psi,\psi^{\ast}]}{\delta \psi_t^{\ast}(x,t)} \, 
   \Bigr ]
   =
   0 \> 
\end{equation}
reproduces Eq.~\eqref{NNLSE}. Note that in the derivation here 
we used
\begin{equation}\label{dpsidpsi}
   \frac{\delta \psi(-x,t)}{\delta \psi(x',t')} 
   =
   \delta(x + x') \, \delta(t - t') \>,
\end{equation}
which provides a factor of $2\,g$ multiplying the nonlocal term.  
The existence of this action formulation immediately leads to the 
fact that the Hamiltonian $H$ given by Eq.~\eqref{Hvalue} is conserved. 

%
%
\section{\label{s:exact}Exact solution}

The exact one-soliton solution to Eq.~\eqref{NNLSE} is given by
\begin{equation}\label{exact}
   \psi(x,t)
   =
   A(\beta,\gamma) \, \sech^{\gamma}(\beta x) \,
      \rme^{\rmi \, \omega \, t}
   \qc
   \gamma=1/\kappa \>,
\end{equation}
provided that
\begin{equation}\label{Adef}
   \omega
   =
   ( \gamma \beta)^2
   \qc
   2 g \, [\, 2 A^{2}(\beta,\gamma) \,]^{1/\gamma}
   =
   \beta^2 \, \gamma ( \gamma + 1 ) \>,
\end{equation}
or, explicitly
\begin{equation}\label{Asol}
   A(\beta,\gamma)
   =
   \frac{1}{\sqrt{2}} \, 
   \Bigl [\, \frac{ \beta^2 \, \gamma (\gamma+1) }{ 2 g } \, \Bigr ]^{\gamma/2} \>,
\end{equation}
with $\beta$ being kept arbitrary. Note that Eq.~\ef{exact} agrees with Eq.~\eqref{YangOneSoliton} 
when $g=1$ and $\gamma = 1$.

%
%
\section{\label{s:conserved} Conservation laws}

It is straightforward to see from Eq.~\eqref{NNLSE} that for any initial 
condition, the mass
\begin{equation}\label{Mdef}
   M
   =
   \int \dd{x} | \psi(x,t) |^2
\end{equation}
is conserved.  For the exact solution of \ef{exact}, $M$ is explicitly 
given by
\begin{equation}\label{Mbetagamma}
   M(\beta,\gamma)
   =
   \frac{1}{2 \beta} \,
   \Bigl [\, 
      \frac{\gamma(\gamma+1) \, \beta^2}{2 g} \, 
   \Bigr ]^{\gamma} c_1(\gamma) \>,
\end{equation}
where $c_1(\gamma)$ is given in Eq.~\eqref{c1def}.
In addition, the energy (or the Hamiltonian) given by Eq.~\eqref{Hvalue} 
is also conserved, and for the solution of Eq.~\eqref{exact}, $E(\beta,\gamma)$ 
is given by
\begin{eqnarray}\label{Eexact}
   E(\beta,\gamma)
   &=
   - M(\beta,\gamma) \, \beta^2 \, \frac{\gamma^2 (2\gamma - 1)}{2 \gamma + 1}
   \\
   &=
   -
   \frac{\beta}{2} \,  \frac{\gamma^2 (2\gamma - 1)}{2 \gamma + 1} \,
   \Bigl [\, \frac{\gamma(\gamma+1) \, \beta^2}{2 g} \, \Bigr ]^{\gamma}
   c_1(\gamma) \>.
   \notag
\end{eqnarray}
It can be discerned from Eq.~\eqref{Eexact} that $E(\beta,\gamma)<0$ for 
$\gamma > 1/2$ or $\kappa < 2$. Moreover, and for $g=1$, the critical mass 
$M^{\ast}$ emanating from Eq.~\eqref{Mbetagamma} at $\kappa = 2$ is
\begin{equation}\label{Mstar}
   M^{\ast}
   =
   \frac{\pi}{4} \sqrt{\frac{3}{2}}
   =
   0.9619123726213981 \>.
\end{equation}

We note in passing that the parity operator has the effect: $\mathcal{P} \psi(x,t) = \psi(-x,t) = \pm \psi(x,t)$
which is satisfied by the exact solution Eq.~\eqref{AblowitzII} (enjoying even parity) to Eq.~\eqref{NNLSE}. Moreover,
there are other conservation laws that are directly obtainable from the equations of motion for $\psi_1$ and $\psi_2$.  
These are the two pseudo-masses 
\begin{eqnarray}\label{M12M21def}
   M_{21} 
   &= 
   \int \dd{x} \psi_2^{\ast}(x,t) \, \psi_1(x,t)
   = 
   \int \dd{x} \psi^{\ast}(-x,t) \, \psi(x,t) \>,
   \\  
   M_{12}
   &= 
   \int \dd{x} \psi_1^{\ast}(x,t) \, \psi_2(x,t) 
   =
   \int \dd{x} \psi^{\ast}(x,t) \, \psi(-x,t)
   \notag \>.
\end{eqnarray}
For the exact soliton solution [cf. Eq.~\eqref{AblowitzII}], these two pseudo-masses 
are equal and also are to the regular mass. However, once we distort the initial state 
from the exact solution, these two conserved quantities are complex conjugates of one 
another.
 
%
%
\section{\label{s:Derrick}Derrick's theorem}

Derrick's theorem~\cite{doi:10.1063/1.1704233} states that a solitary 
wave solution of a Hamiltonian dynamical system is stable, if it is 
stable to scale transformations of the form: $x \mapsto \alpha x$ ($\alpha>0$) 
when we keep the mass of the solitary wave fixed. Let us consider solitary 
wave functions of the form
\begin{equation}\label{solwave}
   \psi(x,t)
   =
   r(x) \,\rme^{- \rmi \omega t } \>,
   \qquad\mathrm{where}\qquad
   r(-x) = r(x) \>,
\end{equation}
which scale as
\begin{equation}\label{scale}
   \psi(x,t) \mapsto \alpha^{1/2} \, r(\alpha x) \,\rme^{- \rmi \omega t } \>,
\end{equation}
and preserve $M$. Moreover, the Hamiltonian $H$ scales as
\begin{subeqnarray}
   H(\alpha) 
   &= 
   H_{\mathrm{kin}}(\alpha) - H_{\mathrm{nl}}(\alpha) \>,
   \\
   H_{\mathrm{kin}}(\alpha)
   &=
   \int \dd{x} | \partial_x \psi(x,t) |^2
   =
   \alpha^2 \int \dd{z} | \partial_z r(z) |^2 > 0 \>,
   \\
   H_{\mathrm{nl}}(\alpha)
   &=
   \frac{g}{\kappa + 1}
   \int \dd{x} 
   [\, |\psi(x,t)|^2 + |\psi(-x,t)|^2 \, ]^{\kappa+1}
   \notag \\
   &=
   \frac{g \, 2^{\kappa+1} \alpha^{\kappa}}{\kappa+1}
   \int \dd{z} [\, r^{\ast}(z) \, r(z)\, ]^{\kappa+1} > 0 \>,
\end{subeqnarray}
and thus we can write it as
\begin{equation}\label{HH1H2}
   H(\alpha) = \alpha^2 \, h_1 - \alpha^{\kappa} \, h_2 \>,
   \qquad
   h_1 > 0 \>, \qquad h_2 > 0 \>.
\end{equation}
It can be shown that the minimum of $H(\alpha)$ is $h_1 = ( \kappa/2 ) \, h_2$,
i.e., upon solving $\pdv{H(\alpha)}{\alpha}\Big|_{\alpha=1}=0$. The second derivative
of $H(\alpha)$ wrt $\alpha$ and evaluated at $\alpha = 1$ gives the stability requirement
in question:
\begin{equation}\label{Dcond}
   \ppdv{H(\alpha)}{\alpha}\Big|_{\alpha=1}
   =
   2 \,(2 - \kappa) \, h_1 \ge 0 \>.
\end{equation}
This result indicates that solutions are unstable to changes in the 
width, compatible with the conserved mass, when $\kappa > 2$.  The case $\kappa=2$ is a marginal case where it is known that blowup occurs at a critical mass \cite{COOPER1993344}. 

The exact solution of Eq.~\ef{exact} has the property that it extremizes 
the Hamiltonian subject to the constraint of fixed mass. For the exact case, 
we find
\begin{subeqnarray}\label{H1andH2eval}
   H_{\mathrm{kin}}
   &=
   \frac{\gamma^2}{2\gamma+1} \, \frac{\beta}{2} \, 
   \Bigl [\, \frac{\gamma (\gamma+1)\beta^2 }{2 \, g} \,\Bigr ]^{\gamma} c_1(\gamma)
   \label{H1-val} \>, \\
   H_{\mathrm{nl}}
   &=
   \frac{2 \gamma^3}{2\gamma+1} \, \frac{\beta}{2} \, 
   \Bigl [\, \frac{\gamma (\gamma+1)\beta^2 }{2 \, g} \,\Bigr ]^{\gamma} c_1(\gamma)\>,
   \label{H2-val} 
\end{subeqnarray}
so that the exact solution is indeed an extremum of the Hamiltonian wrt 
scale transformations, with $H_1 = (\kappa/2) \, H_2$.

%
%
\subsection{\label{ss:VKmethod}Vakhitov-Kolokokov stability criterion}

In the case of the NLSE, one can perform a linear stability analysis of 
the exact solutions, that is, by letting
\begin{equation}\label{nlse.e:1}
   \psi(x,t) 
   = 
   [\,  \psi_\omega(x) + r(x,t) \, ] \, e^{-i \omega t} \>.
\end{equation}
To first order in (the small in amplitude) $r(x,t)$, we arrive at 
\begin{equation}\label{nlse.e:2}
   \partial_t \, r(x,t)
   =
   A_{\omega} \, r(x,t) \>,
\end{equation}
and this way, one can study the eigenvalues of the differential operator 
$A_\omega$. If the spectrum of $A_\omega$ is imaginary, then the solutions 
are spectrally stable. Vakhitov and Kolokokov (V-K) showed in~\cite{Vakhitov:1973aa} 
that when the spectrum is purely imaginary, then $\dd{M(\omega)}/\dd{\omega} < 0$
holds. Alongside, they showed that when
\begin{equation}\label{VK}
   \dd{M(\omega)}/\dd{\omega} > 0 \>,
\end{equation}
there is a real positive eigenvalue, thus rendering the solution to be 
linearly unstable. Assuming that the same argument holds for the NNLSE, 
we have the following result from the V-K criterion [cf.~Eq.~\ef{VK}]:
For the solutions of Eq.~\ef{exact} with $\omega = - \gamma^2 \beta^2$, 
we find from \ef{Mbetagamma} that
\begin{equation}\label{MassOmega}
   \fl
   M(\omega,\gamma)
   =
   \frac{\beta^{2 \gamma - 1}}{2} \, 
   \Bigl [\, \frac{\gamma(\gamma+1)}{2 g} \, \Bigr ]^{\gamma}
   c_1(\gamma) 
   =
   \frac{(- \omega)^{\gamma - 1/2} \, \gamma \,}{2} 
   \Bigl [\, \frac{(\gamma+1)}{2 \gamma g} \, \Bigr ]^{\gamma} \> 
   c_1(\gamma),   
\end{equation}
so that
\begin{equation}\label{dMdomega}
   \fl
   \pdv{M(\omega,\gamma)}{\omega}
   =\frac{1}{2}
   (1/2 - \gamma) \,
   (- \omega)^{\gamma - 3/2} \, \gamma \, 
   \Bigl [\, \frac{(\gamma+1)}{2 \gamma g} \, \Bigr ]^{\gamma}
   c_1(\gamma)
   \qc
   \omega < 0 \>. 
\end{equation}
Thus for $\gamma < 1/2$ or $\kappa > 2$ the solitary waves are unstable. 
This agrees with the result of Derrick's theorem. 


%
%
\section{\label{s:2CC}2CC variational \ansatz}

The NNLSE of Eq.~\ef{NNLSE} is invariant under the parity transformation 
$x \rightarrow -x$, and since the exact solution~\ef{exact} is even under 
parity, we choose a variational \ansatz\ of the form:
\begin{equation}\label{VarAnsatz}
   \tpsi[\, x,Q(t)\,]
   =
   A(t) \, \sech^{\gamma}[\, x / G(t) \,] \,
   \exp[\, \rmi \, ( - \theta(t) + \Lambda(t) \, x^2 ) \,] \>.
\end{equation}
We note that Eq.~\eqref{VarAnsatz} is also even under parity, i.e., 
$\tpsi[\, -x,Q(t)\,] = \tpsi[\, x,Q(t)\,]$. This way, the conserved 
mass is given by
\begin{equation}\label{Mvar}
   M
   =
   \int \dd{x} \, | \tpsi(x,Q(t)) |^2
   = 
   G(t) \, A^2(t) \, c_1(\gamma) \>,
\end{equation}
where $c_1(\gamma)$ is given in Eq.~\ef{c1def}. So $A(t)$ and $G(t)$ 
are not independent variables, and we can set $A^2(t) = M/[\, G(t) \, c_1(\gamma) \,]$.
Since the phase $\theta(t)$ does not enter in the dynamics, it
is ignored, thus leaving two independent CCs:
\begin{equation}\label{Qvalues}
   Q^{\mu}(t) = \{\, G(t), \Lambda(t) \,\} \>.
\end{equation}
We choose initial conditions so as to agree with the exact solution of 
Eq.~\eqref{exact} apart from a small perturbation of $\Lambda(t)$ at $t=0$:
\begin{equation}\label{initial0}
   G(0) = G_0
   \qc
   \Lambda(0) = \Lambda_0
   \qc
   A(0) = A_0 
   =
   \frac{1}{\sqrt{2}} \, 
   \Bigl [\, 
      \frac{ \gamma (\gamma+1) }{ 2 \, g \, G_0^2 } \, 
   \Bigr ]^{\gamma/2} \>, 
\end{equation}
with $G_0$ being arbitrary and $\Lambda_0 = 0.01$. This way, 
and as per the above initial conditions for the 2CC ans\"atz, 
the mass is the soliton mass given by:
\begin{equation}\label{M2CC}
   M
   =
   G_0 \, A^2_0 \, c_1(\gamma)
   =
   \Bigl [\, 
      \frac{ \gamma (\gamma+1) }{ 2 \, g \, G_0^2} \, 
   \Bigr ]^{\gamma} \, \frac{G_0 \, c_1(\gamma)}{2} \>, 
\end{equation}
and thus the value of $G_0$ fixes the mass. We usually set $G_0 = 1$.

\begin{figure}[t]
\centering
\includegraphics[width=0.95\columnwidth]{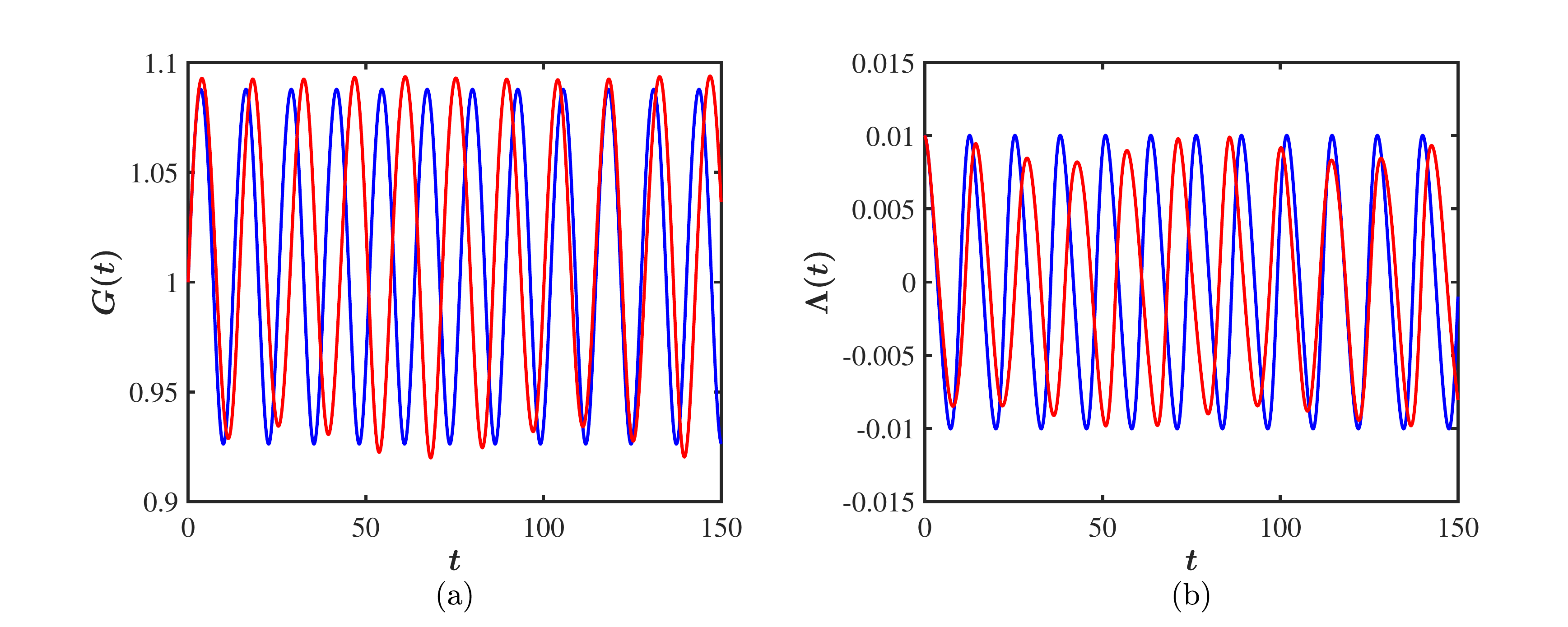}
\caption{\label{fig:stable} 
The temporal evolution of $G(t)$ and $\Lambda(t)$ for the 2CC 
case when $\kappa = 3/2$, with $G_0 = 1$ and $\Lambda_0 = 0.01$. 
The solid blue and red lines correspond to results obtained from 
the 2CC approximation [cf.~Eqs.~\ef{EOM-I}] and numerical simulation 
of the NNLSE.}
\end{figure}

The Lagrangian for the 2CC variational \ansatz\ of Eq.~\ef{VarAnsatz}
is given by
\begin{equation}
L[Q,\dot Q] = T[Q,\dot Q]-H[Q] \>, 
\end{equation}
where
\begin{equation}
T[Q,\dot Q ]=  - M G^2  \dot{\Lambda} \, \frac{c_2(\gamma)}{c_1(\gamma)} \>, 
\end{equation}
\begin{eqnarray} \label{Hamiltonian_V}
   \fl
   H[\,Q\,]
   &=
   M \,
   \Bigl \{\,
     4 G^2  \, \Lambda^2 \, \, 
      \frac{c_2(\gamma)}{c_1(\gamma)}
      +
      \frac{\gamma^2}{2 \gamma + 1} \, \frac{1}{G^2} 
      -
      \frac{2 \, \gamma^3}{2 \gamma + 1} \, 
      \Bigl ( \frac{G_0}{G} \Bigr )^{1/\gamma} \frac{1}{G_0^2} \,
   \Bigr \} \>,
\end{eqnarray}
where $c_2(\gamma)$ is given in Eq.~\eqref{c2def}. 
This way, the equations of motion for the variational parameters 
are then given by:
\begin{subeqnarray}\label{EOM-I}
   \dot{G}
   &=
   4 \, G \, \Lambda \>,
   \label{EOM-G} \\
   \dot{\Lambda}
   &=
   - 
   4 \, \Lambda^2
   +
   \frac{1}{G^4} \,
   \Bigl \{\,
      1
      -
      \Bigl ( \frac{G_0}{G} \Bigr )^{1/\gamma -2} \,
   \Bigr \} \,
   \frac{\gamma^2}{2 \gamma + 1} \, \frac{c_1(\gamma)}{c_2(\gamma)}  \>.
   \label{EOM-Lambda}
\end{subeqnarray}
In Fig.~\ref{fig:stable}, we present results for the case with $\kappa = 3/2$, 
$G_0 = 1$, and $\Lambda_0 = 0.01$. In particular, the left and right panels
showcase the temporal evolution of $G(t)$ and $\Lambda(t)$, respectively, where
the solid blue and red lines correspond to results of the variational approximation
and NNLSE with initial conditions of $\psi(x,0) = \tpsi(x,0)$, for comparison. The 
numerical solutions were obtained by using MATLAB's \texttt{ODE113} integrator which 
is a variable-step, variable-order (VSVO) Adams-Bashforth method (see~\cite{ode113_ref}, 
and references therein). We solve the NNLSE
on $x \in [-L,L]$ with $L = 10$ numerically, supplemented by Dirichlet boundary conditions. 
As per the spatial discretization, we considered a centered yet fourth-order accurate
finite difference approximation. We further tested our numerical simulation results by 
considering other integrators, spatial discretizations, and boundary conditions, such 
as the \texttt{ETDRK4} integrator and Fourier spectral collocation~\cite{KassamTrefethen}, 
and we obtained essentially similar results. For the 2CC approximation (shown in blue in 
the figure), $G(t)$ and $\Lambda(t)$ oscillate, indicating stability and agreeing reasonably 
well with numerical solutions of the NNLSE (shown in red in the figure).

%
%
%
%
%
%
The conserved energy in the 2CC approximation is expressed as a function
of $G(t)$ and $\Lambda(t)$ as follows:
\begin{equation}\label{E2CC}
   \fl
   E(G,\Lambda,G_0,\gamma)
   =
   M(G_0,\gamma)  \,
   \Bigl \{\,
      4 \, G^2 \, \Lambda^2 \, \frac{c_2(\gamma)}{c_1(\gamma)}
      +
      \frac{1}{G^2} \,
      \Bigl [\,
         1  
         -
         2 \, \gamma \, 
         \Bigl ( \frac{G_0}{G} \Bigr )^{1/\gamma - 2} \,
      \Bigr ] \,
      \frac{\gamma^2}{2 \gamma + 1} \, 
   \Bigr \} \>.
\end{equation}
The energy of the perturbed solution is given by its value at $t=0$
\begin{equation}\label{Et0}
   \fl
   E(G_0,\Lambda_0,G_0,\gamma)
   =
   M(G_0,\gamma) \,
   \Bigl \{\,
      4 \, G_0^2 \, \Lambda_0^2 \, \frac{c_2(\gamma)}{c_1(\gamma)}
      -
      \frac{1}{G_0^2} \, 
      \frac{\gamma^2 (2 \gamma - 1)}{2 \gamma + 1} \,
   \Bigr \} \>.
\end{equation}
$\Lambda(t)$ oscillates about zero, and when it is at zero, $G$ reaches its maximum value $G_m$ which is determined from its initial energy so that 
the maximum value that $G(t)$ can have is when $\Lambda(t) = 0$, or when 
\begin{equation}\label{EGmax}
   \fl
   E(G_m,0,G_0,\gamma)
   =
   M(G_0,\gamma)  \,
   \Bigl \{\,
      \frac{1}{G_m^2} \,
      \Bigl [\,
         1  
         -
         2 \, \gamma \, 
         \Bigl ( \frac{G_0}{G_m} \Bigr )^{1/\gamma - 2} \,
      \Bigr ] \,
      \frac{\gamma^2}{2 \gamma + 1} \, 
   \Bigr \} \>.
\end{equation}
The maximum value of $G$ for $\kappa=3/2$ when we choose $\Lambda_0=0.01, G_0=1$ 
is given by $ G_{m} = 1.087874$. This value of $G_{m}$ is seen in both the 
2CC approximation and in the numerical simulations as seen in Fig.~\ref{fig:stable}. 

\begin{figure}[t]
\centering
\includegraphics[width=0.95\columnwidth]{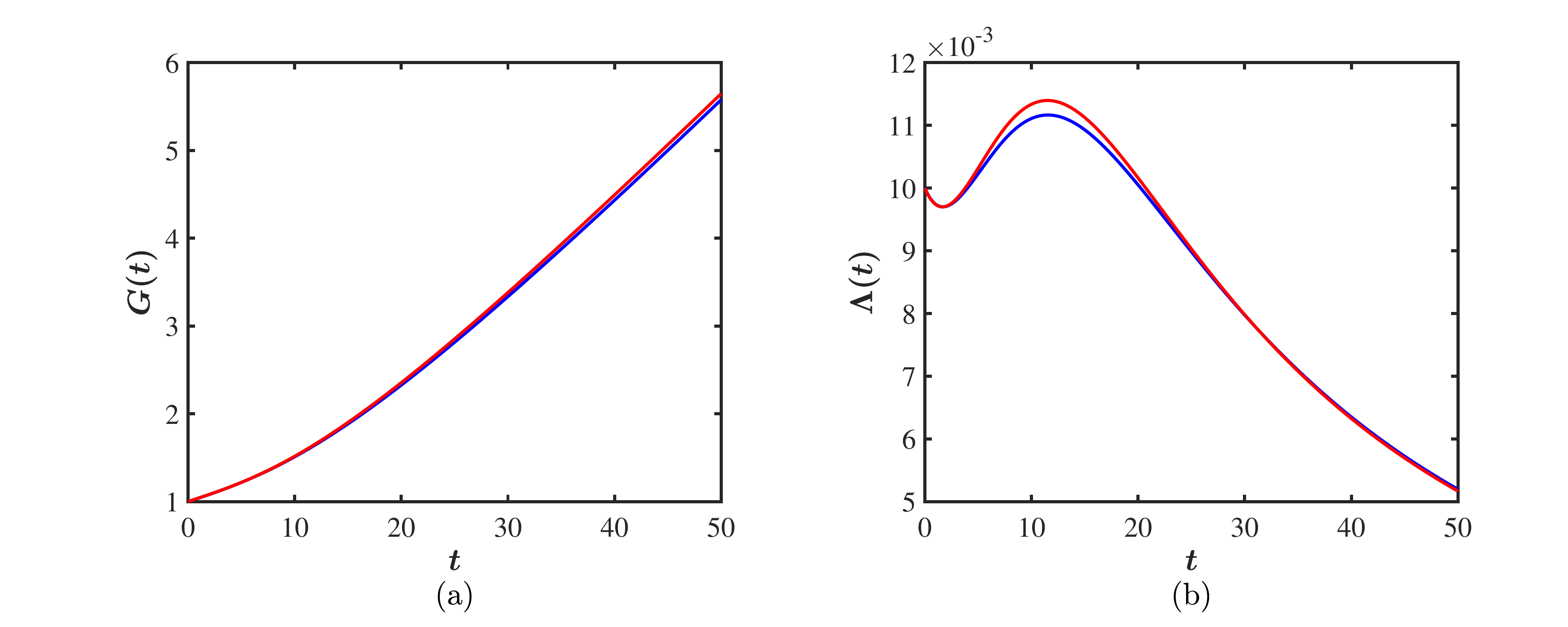}
\includegraphics[width=0.95\columnwidth]{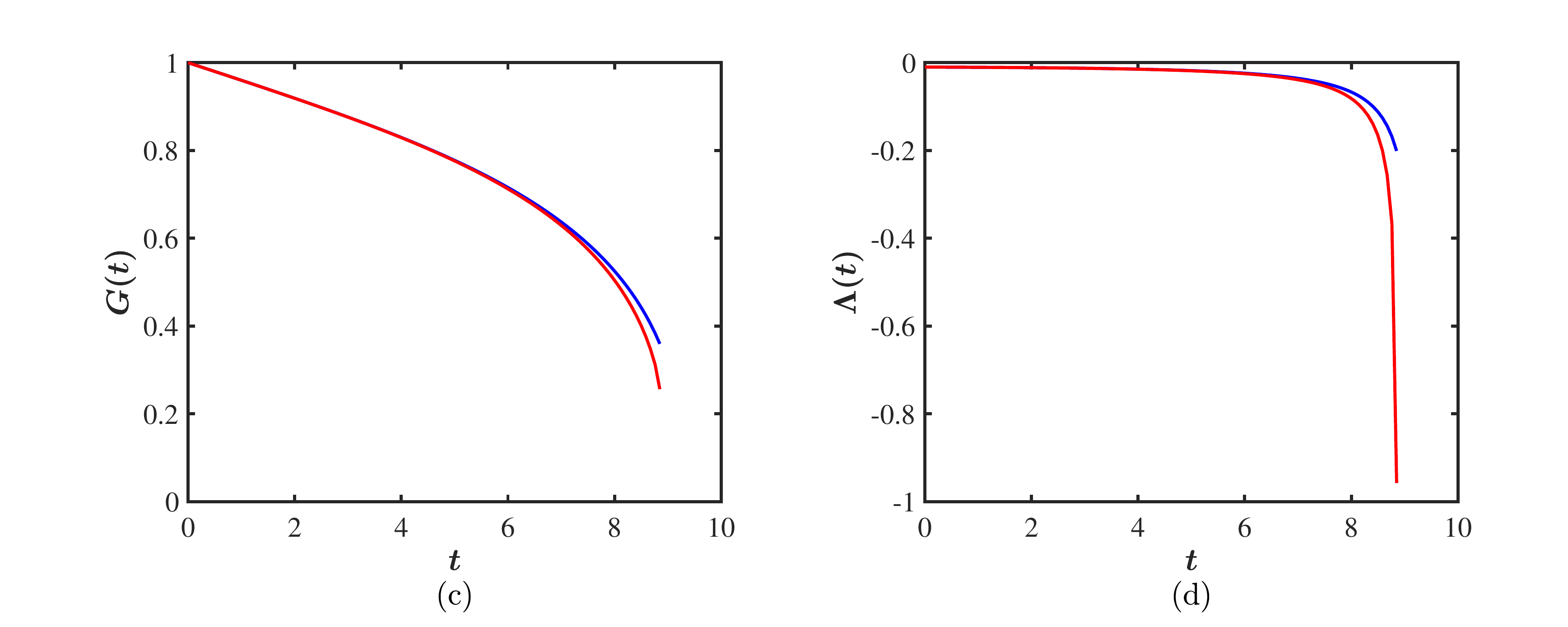}
\caption{\label{fig:unstable}
Same as Fig.~\ref{fig:stable} but with $\kappa=2.2$ (top panels)
and $\kappa=2.1$ (bottom panels), respectively ($G_0 = 1$ in both cases). 
In particular, the temporal evolution of $G(t)$ and $\Lambda(t)$ 
for the 2CC and NNLSE is presented for $\Lambda_0 = 0.01$ (blowup)
and $\Lambda_0 = -0.01$ (collapse) in the top and bottom panels,
respectively. Similarly, the blue solid line corresponds to the 
solution of the 2CC approximation [cf.~Eqs.~\ef{EOM-I}] whereas
the solid red line to the numerical solution of the NNLSE. 
}
\end{figure}

The cases when $\kappa = 2.2$ and $\kappa=2.1$ are shown in the top and bottom panels
of Fig.~\ref{fig:unstable}. Specifically, the onset of collapse is presented in the 
top panels in the figure whereas a case corresponding to blowup is shown at the bottom 
panels therein. We should note that the numerical solutions agree quite well with the 
2CC simulations (shown in red and blue, respectively). 

%
%
%

%
%
\subsection{\label{s:LinerResponse}Linear response}

We now discuss the first-order linear response to Eq.~\ef{EOM-I}. To 
that end, we first set
\begin{eqnarray}\label{LRinitial}
   G(t) = G_0 + \Delta G
   \qc
   \Lambda(t) = \Delta\Lambda(t) \>,
\end{eqnarray}
and substitute them into Eq.~\ef{EOM-I} subsequently. Upon keeping
only first-order terms, we arrive at
\begin{subeqnarray}\label{LR-I}
   \Delta\dot{G}
   &=
   4 \, G_0 \, \Delta\Lambda \>,
   \label{LRG} \\
   \Delta\dot{\Lambda}
   &=
   -
   \frac{ \gamma \, ( 2 \gamma - 1 ) }{ 2 \gamma + 1 } \, 
   \frac{c_1(\gamma)}{c_2(\gamma)} \, \frac{\Delta G}{G_0^5} \>.
   \label{LRLambda}
\end{subeqnarray}
Differentiating Eq.~\eqref{LRG} wrt $t$, and replacing $\Delta\dot{\Lambda}$
therein with Eq.~\eqref{LRLambda}, we find
\begin{equation}\label{LRGequ}
   \Delta \ddot{G} + \omega_G^2 \, \Delta G = 0 \>, 
\end{equation}
where
\begin{equation}\label{}
   \omega_G^2 
   =
   \frac{4 \, \gamma \, ( 2 \gamma - 1)}{2 \gamma + 1} \, 
   \frac{c_1(\gamma)}{c_2(\gamma)} \, \frac{1}{G_0^4}\>.
\end{equation}
Based on the above, the system is unstable wrt width 
$G(t)$ if $\gamma < 1/2$ or $\kappa > 2$, a finding that is independent 
of the value of $G_0$ and in agreement with Derrick's theorem. For 
$G_0 = 1$ and $\kappa = 3/2$, the oscillation period $T_{G} = 2 \pi / \omega_G = 12.6$ 
is in agreement with the stable $G(t)$ oscillations shown in Fig.~\ref{fig:stable}. 
As per the unstable case with $\kappa = 2.1$, the lifetime $\tau_G = 1/|\omega_G| = 7.6$ 
is in reasonable agreement with the lifetime shown in Fig.~\ref{fig:unstable}.

%
%
\subsection{\label{s:blowup}Blowup for $\kappa \ge 2$}

Within the 2CC approach, we can look at the equation of motion for the width 
parameter $G(t)$ in order to see if it goes to zero (blowup) or infinity (collapse). 
It is easiest to look at the energy conservation equation to study these phenomena.  
If we fix $G_0 = 1$, then from energy conservation and the equation of motion for 
$G(t)$, the scaled energy of Eq.~\ef{Lagrangian} can be rewritten as:
\begin{equation}
   \frac{E}{M} 
   =
   \frac {{\dot G}^2}{4} \frac{c_2(\gamma)}{c_1(\gamma)}
   +
   \frac{1}{G^2} \, \frac{\gamma^2}{2 \gamma + 1}
   -
   \frac{2 \, \gamma^3}{G^{1/\gamma} (2 \gamma + 1)}.
\end{equation}
We immediately notice that when $\kappa=2$, the last two terms 
cancel so that for the exact solution $\dot{G}(t)$ is constant. To 
determine a critical mass, one needs to allow for initial conditions 
with mass greater than the exact solution, and assume a self-similar 
shape. To this end, we proceed as follows. Let us look first at the 
case with $\kappa > 2$ where it is possible to have the width parameter 
$G(t) \rightarrow 0$. Upon rewriting the energy equation as
\begin{equation}
   \dot G^2 
   = 
   4 \, \frac{c_1(\gamma)}{c_2(\gamma)}  
   \left( 
      \frac{E}{M} 
      +  
      \frac{2 \, \gamma^3}{G^{1/\gamma} (2 \gamma + 1)}
      - 
      \frac{1}{G^2} \, \frac{\gamma^2}{2 \gamma + 1} 
   \right),
\end{equation}
and considering $G \rightarrow 0$, we can ignore the first and last terms,
thus obtaining 
\begin{equation}\label{Geqblowup}
   \dot G 
   =  
   -  
   \sqrt { \frac{c_1(\gamma)}{c_2(\gamma)}   
   \frac{2 \, \gamma^3}{G^{1/\gamma} (2 \gamma + 1)} }
\end{equation}
for blowup. In other words, when we get near the critical time $t^{\ast}$,
we have
\begin{equation}
   G(t) \propto (t-t^{\ast})^{2/(\kappa+2)} \, \>. 
\end{equation}
To explore the critical mass, we do not use the exact solution as our 
starting point. Instead, we rewrite the nonlinear term in terms of the 
mass $M$ of the solitary wave via
\begin{equation}
   A^2(t) \rightarrow  M \beta(t)/c_1(\gamma).
\end{equation}
This way, when $q(t)=0$, we have that 
\begin{eqnarray}
   \fl
   \tilde{H}_{\mathrm{nl}}
   &=
   \frac{g}{\kappa + 1} 
   \tint \dd{x} [\, | \tpsi(x,t) |^2 +  | \tpsi(-x,t) |^2 \,]^{\kappa+1} 
   \label{nnltermI} \\
   \fl
   &=
   \frac{g \, \gamma }{\gamma + 1} \, A^{2/\gamma + 2}
   G \tint \dd{y} [\,2  \sech^{2 \gamma}(y)  \,]^{1/\gamma+1} 
   = 
   M \frac{2 \gamma^2 }{2 \gamma+1}
   \frac{2 g}{\gamma+1}  
   \Bigl ( 
      \frac{ 2M}{G \, c_1(\gamma)} 
   \Bigr )^{1/\gamma} \>.
   \notag  
\end{eqnarray}
For blowup to set in when $\gamma=1/2$, this term must be bigger than the 
negative $1/G^2$ term in the energy, so for $g=1$, one gets the condition:
\begin{equation}\label{Mstar-II}
   M 
   \ge
   M^{\ast}
   =
   \frac{\pi}{4} \sqrt{\frac{3}{2}}
   =
   0.9619123726213981 \>,
\end{equation}
which in turn is in agreement with Eq.~\ef{Mstar}.  

%
%
\section{\label{s:4CC}4CC variational \ansatz}

When we displace the position, we destroy the exact parity symmetry 
of the solutions of the NNLSE. In this section, we study this by 
choosing a parity-breaking, four collective coordinate (4CC) \ansatz\ 
for our variational wave function of the form:
\begin{subeqnarray}\label{4CCAnsatz}
   \tpsi(x,Q(t))
   &=
   A(t) \, \sech^{\gamma}[\,(x - q(t))/G(t) \,] \, 
   \rme^{\rmi \phi(x)} \>,
   \label{varpsi4CC} \\
   \phi(x,Q(t))
   &=
   - \theta(t) + p(t) \,(x - q(t)) + \Lambda(t) \, (x - q(t))^2 \>.
   \label{varphi4CC}
\end{subeqnarray}
The mass is again given by $M = G(t) \, A^2(t) \, c_1(\gamma)$, so that 
$A(t)$ can be eliminated in favor of $G(t)$ as an independent variable.
Similar to the 2CC case, the phase $\theta(t)$ in the 4CC case does not 
enter in the dynamics so it can be ignored. Thus, we are left with four 
variational parameters:
\begin{equation}\label{Q4CC}
   Q(t) = \{\, q(t), p(t), G(t), \Lambda(t) \,\} \>.
\end{equation}
The Lagrangian in this case is given by
\begin{equation}\label{Lagrangian4CC}
   L[Q,\dot{Q}]
   =
   T[Q,\dot{Q}] - H[Q] \>,
\end{equation}
where
\begin{subeqnarray}\label{THdefs}
   \fl
   T[Q,\dot{Q}]
   &=
   M \,
   \Bigl \{\,
      p \, \dot{q}
      -
      G^2 \, \dot{\Lambda} \, \frac{c_2(\gamma)}{c_1(\gamma)} \,
   \Bigr \}\>,
   \label{HTdef} \\
   \fl
   H[Q]
   &=
   M \, 
   \Bigl \{\,
      p^2
      +
      4 \, G^2 \Lambda^2 \, \frac{c_2(\gamma)}{c_1(\gamma)}
      +
      \frac{1}{G^2} \,
      \Bigl [\, 
         1 
         -
         2 \, \gamma \, 
         \Bigl ( \frac{G_0}{G} \Bigr )^{1/\gamma - 2 } \,
         \!\!\! f(q/G,\gamma) \,
      \Bigr ] \,
      \frac{\gamma^2}{2 \gamma + 1} \,
   \Bigr \} \>,
   \label{HHdef} 
\end{subeqnarray}
together with
\begin{equation}\label{fdef}
   \fl
   f(z,\gamma)
   =
   \frac{2 \gamma +1}{2^{1/\gamma + 2} \, \gamma \, c_1(\gamma)} 
   \tint \dd{y} [\, \sech^{2 \gamma}(y - z) 
   +  
   \sech^{2 \gamma}(y + z) \,]^{1/\gamma+1} \>.
\end{equation}
Thus, the equations of motion are given by:
\begin{subeqnarray}\label{EOM4CC}
   \fl
   \dot{q}
   &=
   2 \, p \>,
   \label{EOMq} \\
   \fl
   \dot{p}
   &=
   \frac{2 \, \gamma^3}{2 \gamma + 1} \, \frac{1}{G^3} \,
   \Bigl (\,
      \frac{G_0}{G}
   \Bigr )^{1/\gamma - 2} \!\! f'(q/G,\gamma)\>,
   \label{EOMp} \\
   \fl
   \dot{G}
   &=
   4 \, G \, \Lambda \>,
   \label{EOMG} \\
   \fl
   \dot{\Lambda}
   &=
   -
   4 \, \Lambda^2 
   +
   \Bigl \{\,
      \frac{1}{G^4} \,
      \Bigl [\,
         1
         -
         \Bigl (\,
            \frac{G_0}{G}
         \Bigr )^{1/\gamma - 2} \!\! f(q/G,\gamma) \,
      \Bigr ] \, 
      \label{EOMLambda} \\
      \fl
      & \hspace{7em}
      -
      \frac{\gamma \, q}{G^5} \,
      \Bigl (\,
            \frac{G_0}{G}
      \Bigr )^{1/\gamma - 2} \!\!f'(q/G,\gamma) \,
   \Bigr \} \, \frac{\gamma^2}{2 \gamma + 1}  \,
   \frac{c_1(\gamma)}{c_2(\gamma)}  \>.
   \notag
\end{subeqnarray}
It should be noted in passing that Eqs.~\eqref{EOM4CC} agree
with the 2CC equations discussed in the previous section when 
$q = p = 0$. 

\begin{figure}[t]
\centering
\includegraphics[width=0.95\columnwidth]{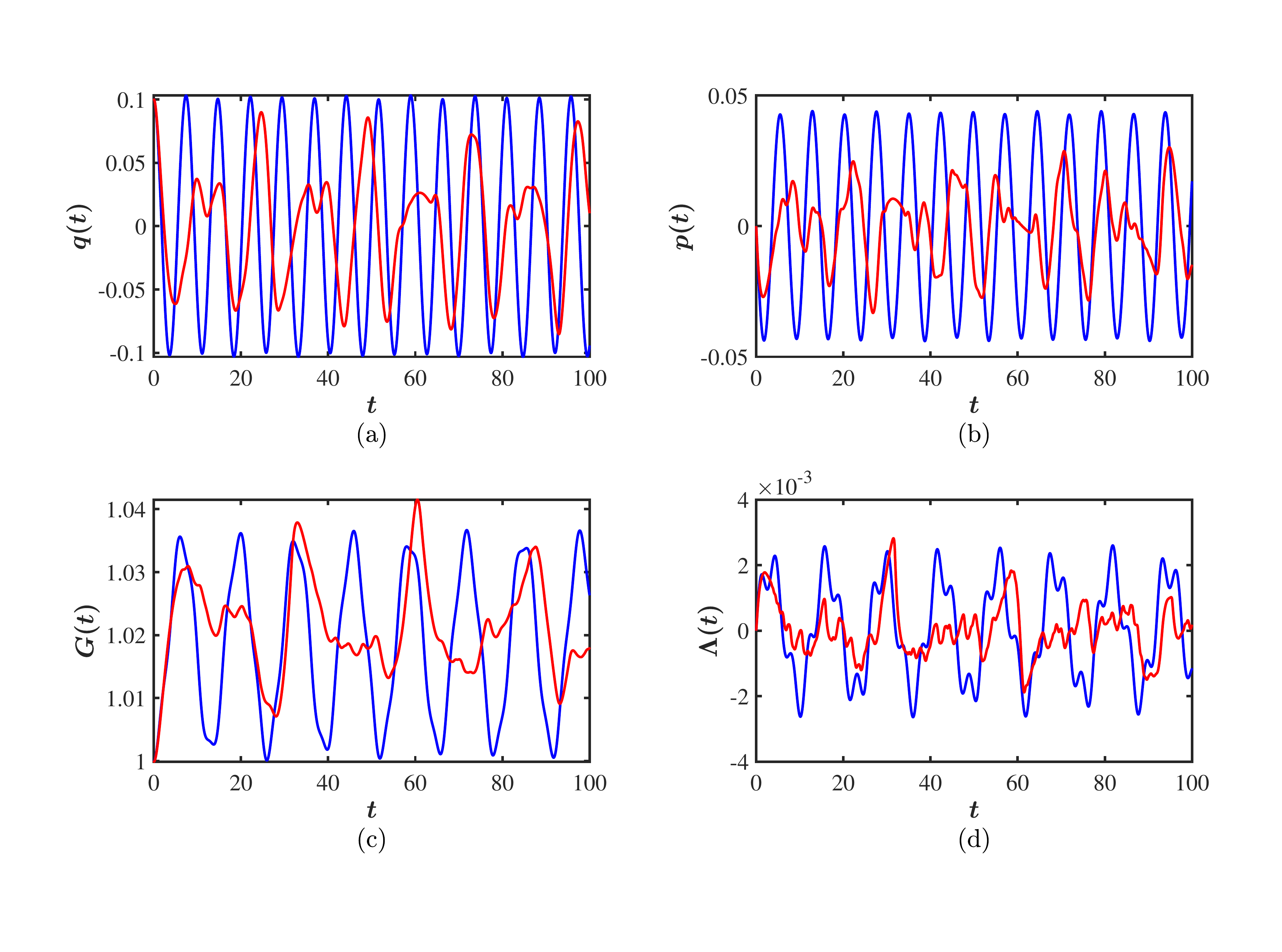}
\caption{\label{f:qpGL4CCSchk32}
Same as Fig.~\ref{fig:stable} but for the 4CC case with 
$\kappa = 3/2$ and $G_0 = 1$. In particular, the temporal
evolution of $q(t)$, $p(t)$, $G(t)$, and $\Lambda(t)$ is
depicted in the panels. Note that results of the 4CC \ansatz\ 
are shown in blue and results for the numerical solution of 
\Schrodinger\,equation are shown in red.}   
\end{figure}

For the discussion that follows next on the 4CC case, we consider 
initial conditions of the form of:
\begin{equation}\label{initial}
   q_0 = 0.1
   \qc
   p_0 = 0
   \qc
   G_0 = 1
   \qc
   \Lambda_0 = 0 \>.
\end{equation}
Typical results for values of $q(t)$, $p(t)$, $G(t)$, and $\Lambda(t)$ are shown 
in Fig.~\ref{f:qpGL4CCSchk32} for the 4CC \ansatz\ along with the numerical findings 
of Eq.~\ef{NNLSE} using the same numerical methods as previously. We note that the 
4CC \ansatz\ does not preserve parity conservation, and also does not preserve the 
conservation of the pseudo-masses $M_{12}$ and $M_{21}$ defined in Eq.~\eqref{M12M21def}. 
For the shifted solution these pseudo-masses are complex. Although they are conserved 
for the exact NNLSE, they are not for the 4CC \ansatz. Because of this shortcoming, we only 
present one such result, namely for $\kappa = 1.5$ and $q_0 = 0.1$. Both the 4CC results 
shown in blue and the numerical solution of the NNLSE shown in red are in substantial 
agreement. Both oscillate about equilibrium values and indicate stability of the soliton.  


%
%
%
%

%
%

%
%
\section{\label{s:NumSim}Numerical simulations of the NNLSE}
\begin{figure}[htp]
\begin{center}
\includegraphics[height=.18\textheight, angle =0]{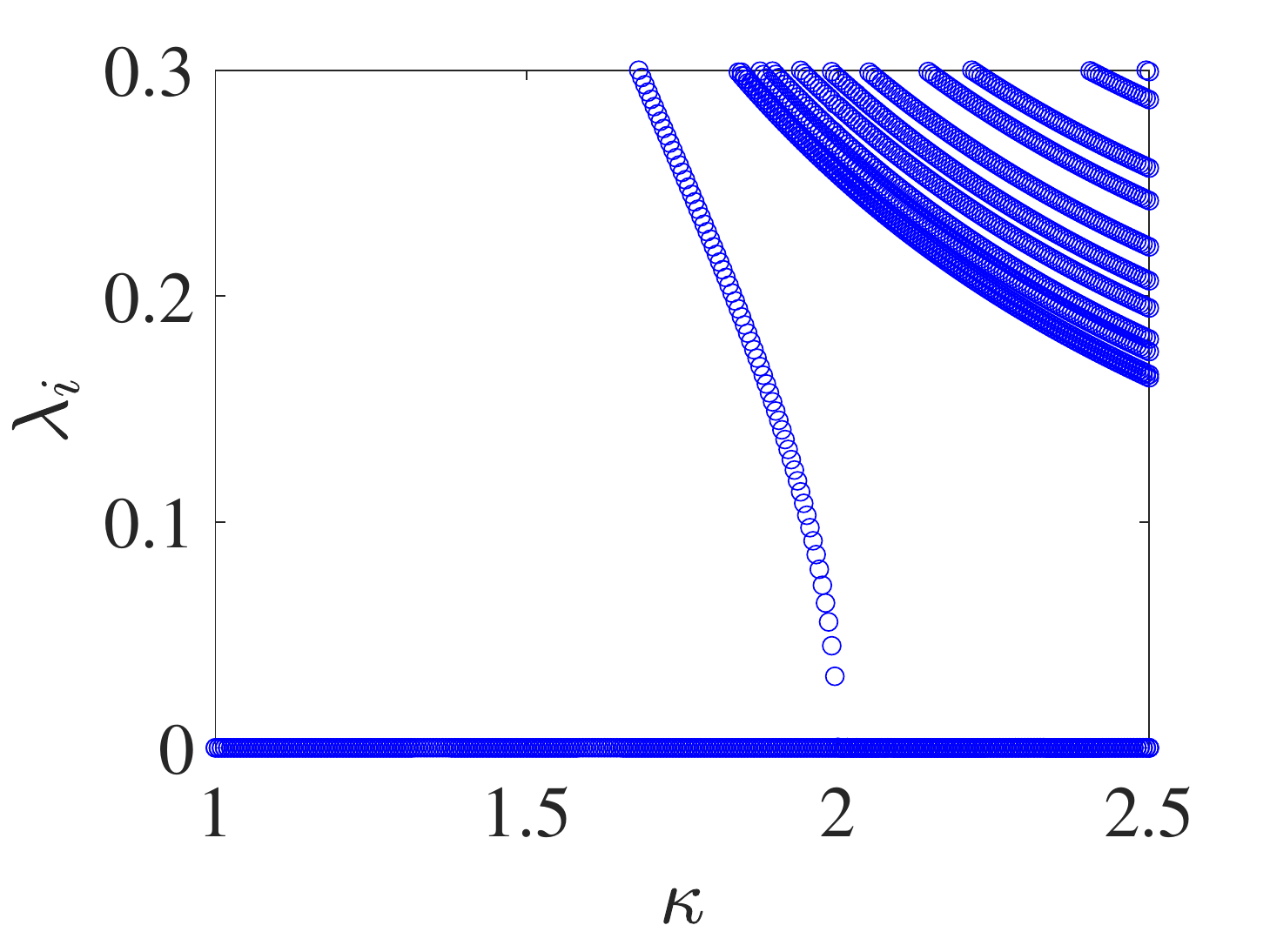}
\includegraphics[height=.18\textheight, angle =0]{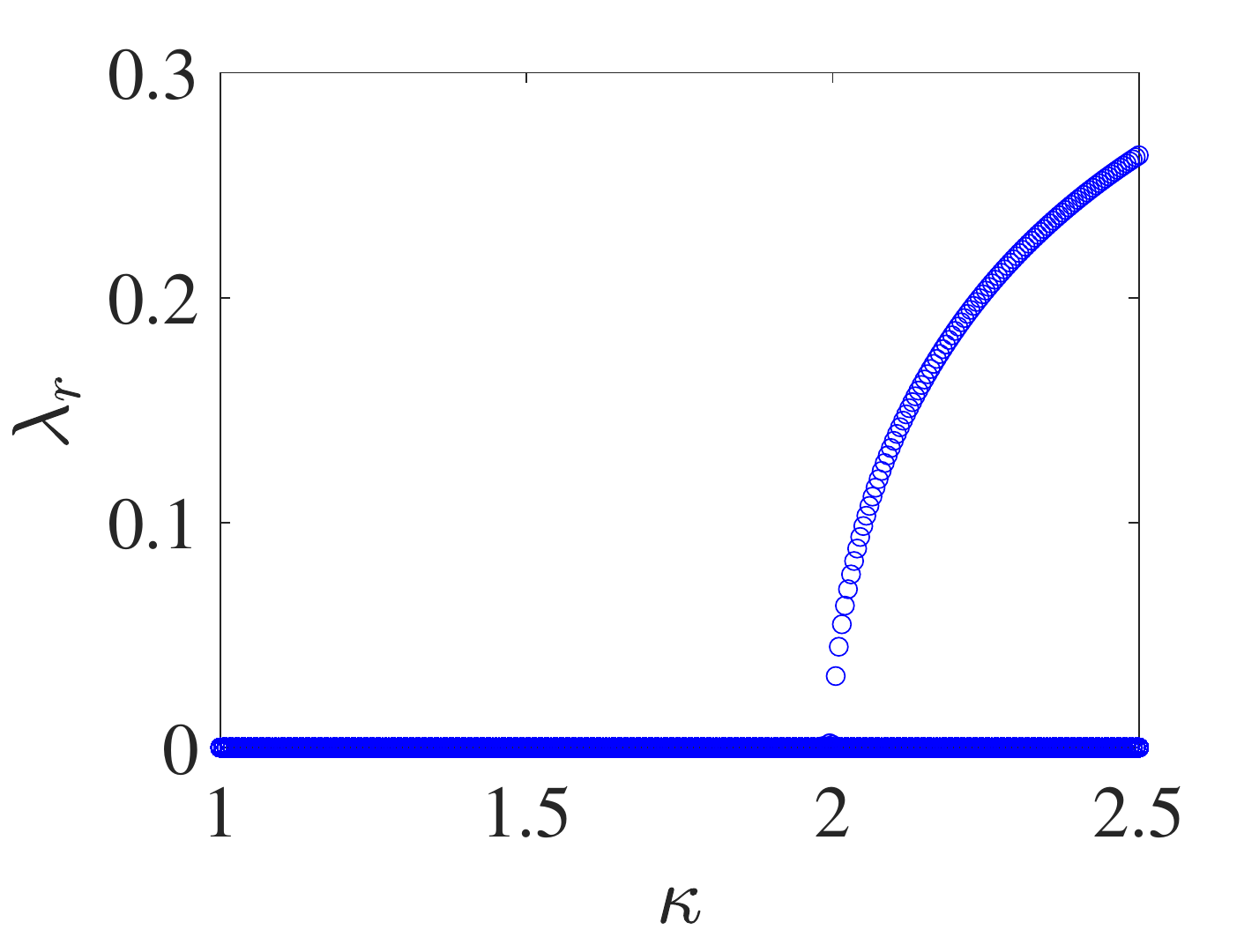}
\end{center}
\caption{The imaginary $\lambda_{i}$ (left) and real $\lambda_{r}$ 
(right) parts of the eigenvalue $\lambda$ as functions of $\kappa$
with $g=1$ and $\beta=1$. At the critical value of $\kappa_{\mathrm{c}}=2$, 
the solution [cf. Eq.~\eqref{exact}] becomes unstable.
}
\label{fig1}
\end{figure}
In this Section, we investigate the stability and spatio-temporal
evolution of the exact solution~\eqref{exact} to the NNLSE~\eqref{NNLSE}.
To do so, we employ the separation of variables ansatz $\psi(x,t)=\psi^{(0)}(x)e^{\rmi \omega t}$,
which is inserted into Eq.~\eqref{NNLSE}, thus obtaining:
\begin{equation}
\frac{\ddd{\psi^{(0)}(x)}}{\dd{x^{2}}}+2g%
\left[|\psi^{(0)}(x)|^{2}+|\psi^{(0)}(-x)|^{2}\right]^{\kappa}\psi^{(0)}(x)-\omega\psi^{(0)}(x)=0.
\label{bvp_stead}
\end{equation}
Then, Eq.~\eqref{bvp_stead} is solved by means of a Newton-Krylov method~\cite{Kelley_book} with Eq.~\eqref{exact} 
as an initial guess. Having identified a stationary solution $\psi^{(0)}(x)$ (upon convergence of
the Newton-Krylov solver), we perform a spectral stability analysis around it using the ansatz:
\begin{equation}
\widetilde{\psi}(x,t)=e^{\rmi \omega t}\left[\psi^{(0)}(x)+%
\varepsilon\left(a(x)e^{\lambda t}+b^{\ast}(x)e^{\lambda^{\ast}t}\right)\right], \quad \varepsilon\ll 1.
\label{stab_ansatz}
\end{equation}
Upon inserting Eq.~\eqref{stab_ansatz} into Eq.~\eqref{NNLSE} we arrive 
(at order $\mathcal{O}(\varepsilon)$) at the eigenvalue problem of the 
form of 
\begin{equation}\label{eig_prob}
\left ( 
\begin{array}{cccc}
  A_{11}        &  A_{12} \\
 -A_{12}^{\ast} & -A_{11}^{\ast}\\
\end{array}
\right )
\left (
\begin{array}{c}
a \\
b \\
\end{array}
\right )
=
\widetilde{\lambda}
\left (
\begin{array}{c}
a \\
b \\
\end{array}
\right ), \quad \widetilde{\lambda} = -\rmi \lambda,
\end{equation}
with matrix elements given by
\begin{subeqnarray}
\fl
   A_{11}
   &=
   \frac{\ddd }{\dd x^{2}} + 2g
   \bigg\{ 
      \kappa 
      \left[
         |\psi^{(0)}(x)|^{2}+|\psi^{(0)}(-x)|^{2}
      \right]^{\kappa-1}
      \left[
         |\psi^{(0)}(x)|^{2}+\psi_{0}(x)\psi_{0}^{\ast}(-x)\mathcal{P}
      \right]
   \notag \\
   \fl
   & \hspace{3em}
   +
   \left[
      |\psi^{(0)}(x)|^{2}+|\psi^{(0)}(-x)|^{2}
   \right]^{\kappa}\bigg\} - \omega \>,
   \notag \\
   \fl
   A_{12}
   &=
   2g\kappa
   \left[
      |\psi^{(0)}(x)|^{2}+|\psi^{(0)}(-x)|^{2}
   \right]^{\kappa-1}
   \left[
      \left(
        \psi^{(0)}(x)
      \right)^{2}
      +
      \psi_{0}(x)\psi_{0}(-x)\mathcal{P}
   \right] \>,
\end{subeqnarray}
where $\mathcal{P}$ stands for the space reflection operator, i.e., 
$\mathcal{P}f(x)=f(-x)$, for a general function $f(x)$.

\begin{figure}[htp]
\begin{center}
\includegraphics[height=.17\textheight, angle =0]{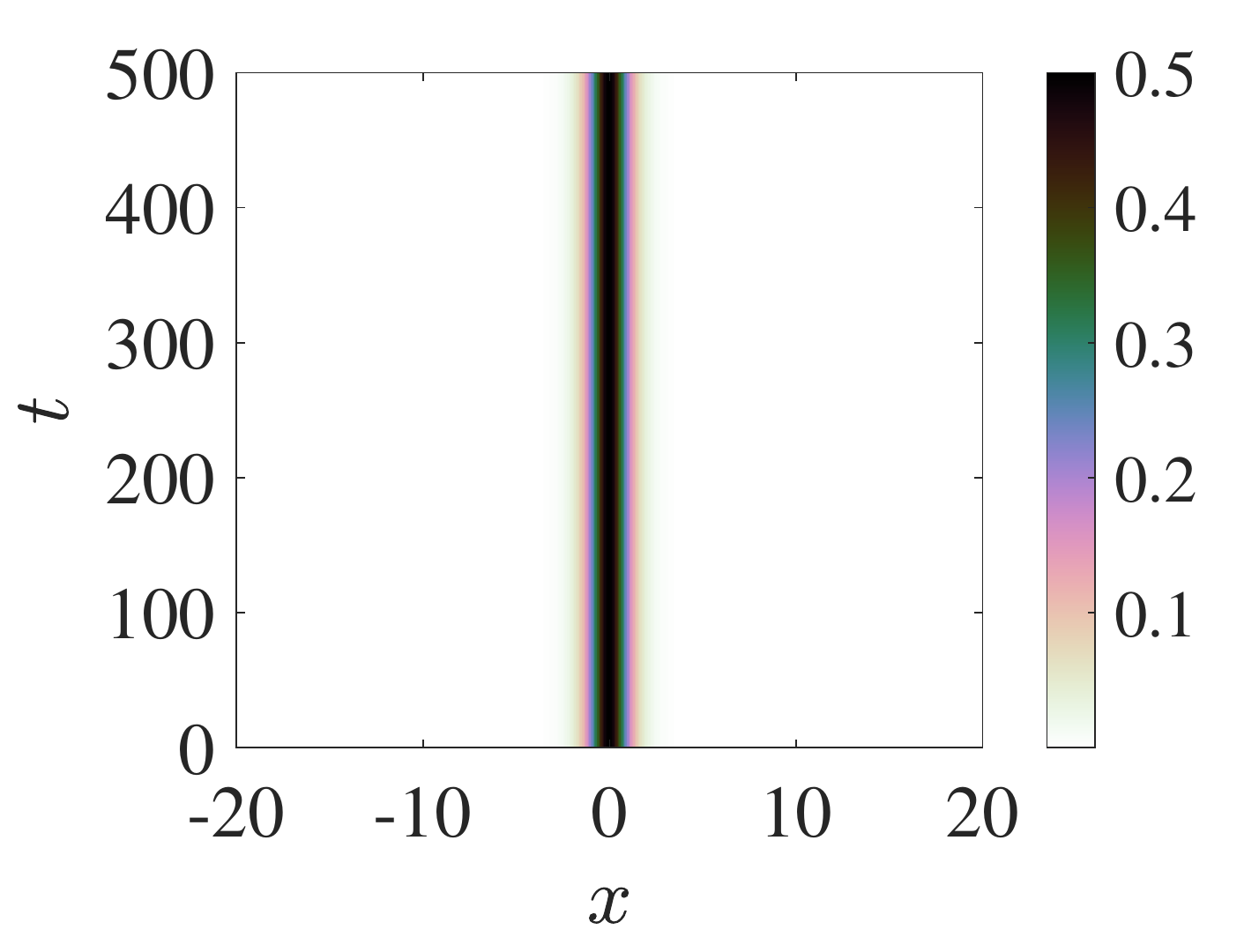}
\includegraphics[height=.17\textheight, angle =0]{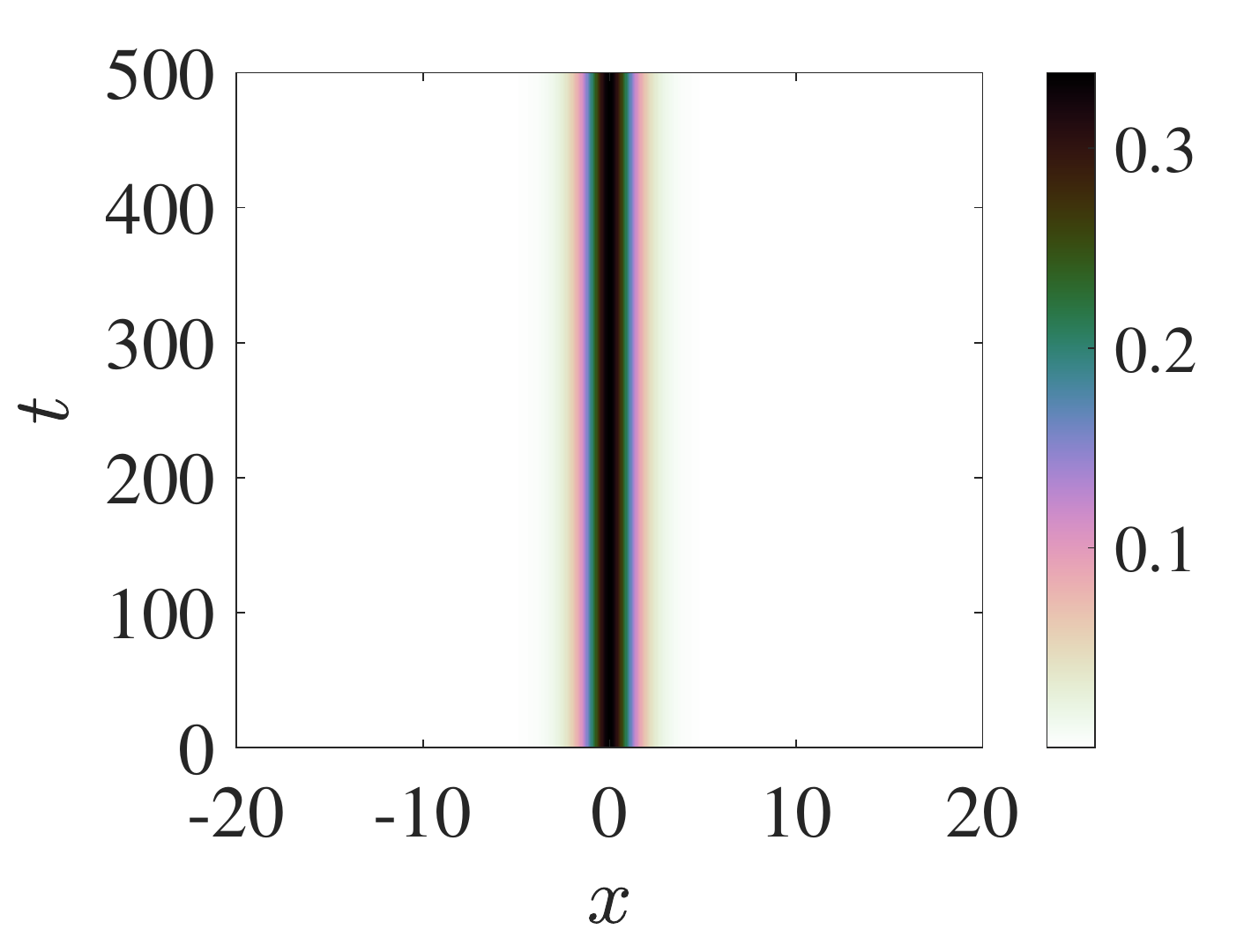}
\includegraphics[height=.17\textheight, angle =0]{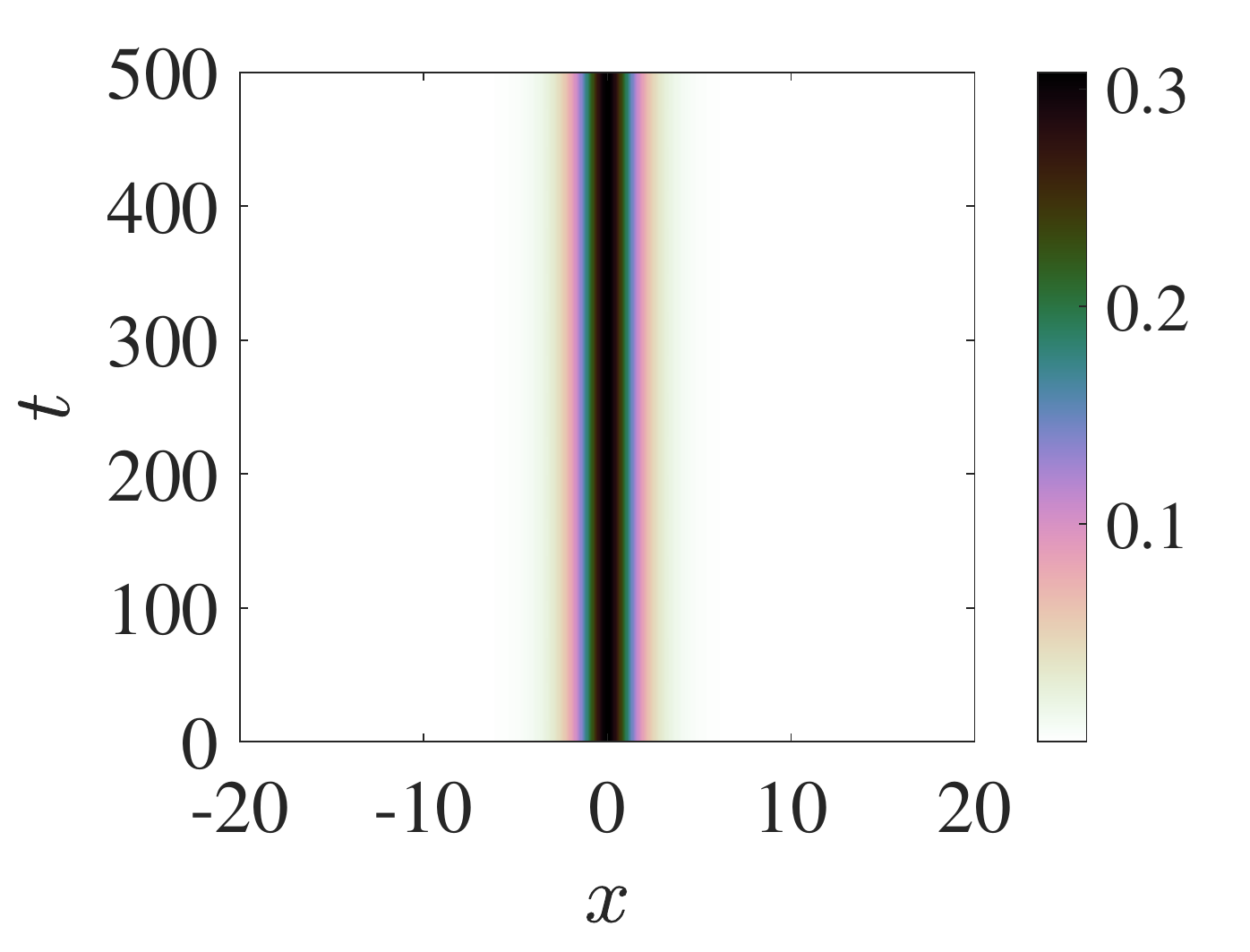}
\includegraphics[height=.17\textheight, angle =0]{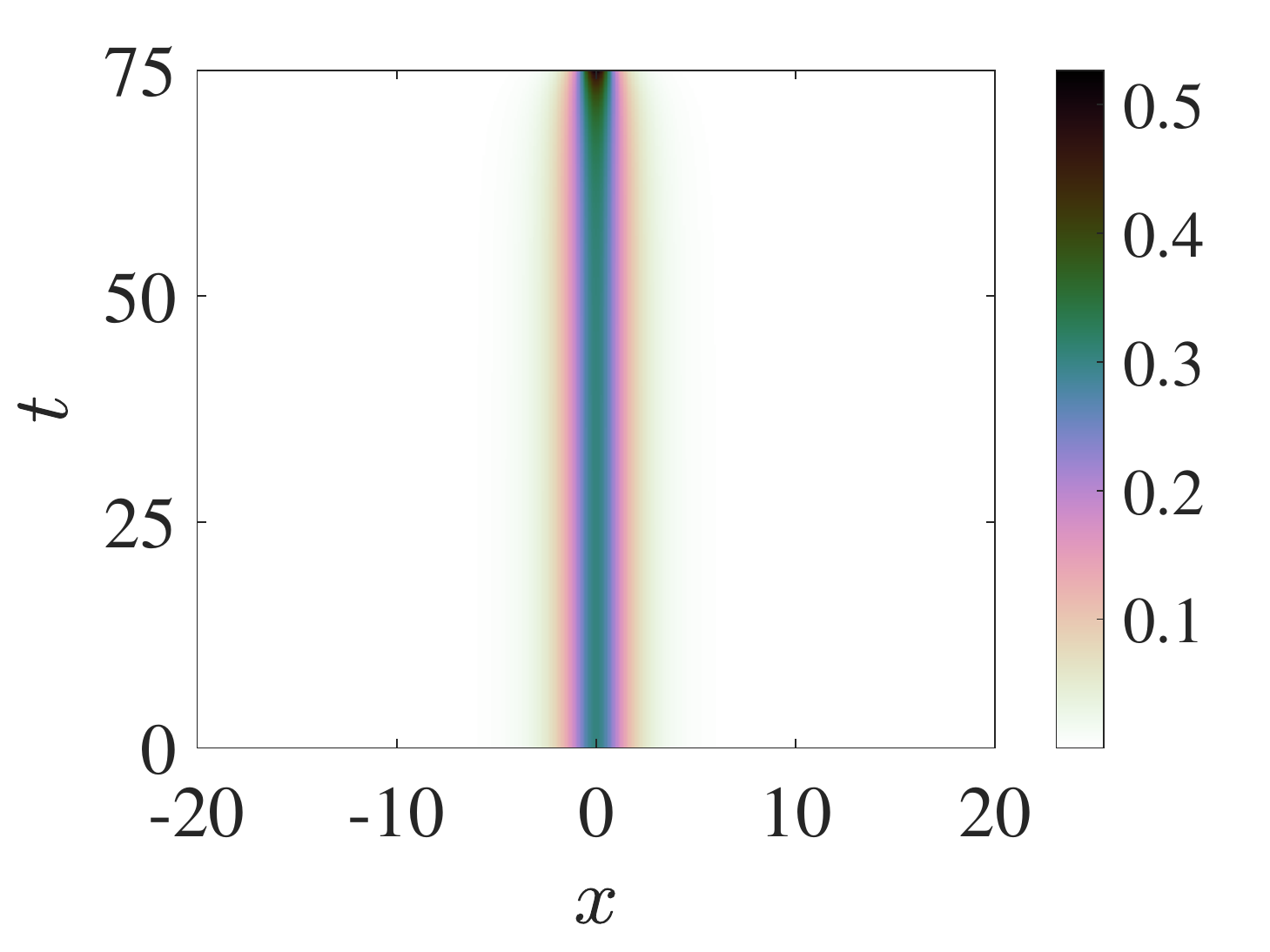}
\end{center}
\caption{Spatio-temporal evolution of the density $|\psi|^{2}$ for values of 
$\kappa$ of: $\kappa=1$ (top left), $\kappa=1.5$ (top right), 
$\kappa=1.95$ (bottom left), and $\kappa=2.1$ (bottom right). 
}
\label{fig2}
\end{figure}

The results of the eigenvalue problem [cf. Eq.~\eqref{eig_prob}] are
shown in Fig.~\ref{fig1} with $g=1$ and $\beta=1$. It can be discerned
from the right panel of the figure, the emergence of a real eigenvalue
at $\kappa_{\mathrm{c}}=2$, thus rendering the solution to be spectrally
unstable (and similar to the local case). Although a detailed study and
understanding of the underlying instability mechanism is of paramount 
importance (see, e.g.,~\cite{Sulem_book} as well as~\cite{Siettos_2003}
and references therein), it is beyond the scope of the present work. Subsequently, 
Fig.~\ref{fig2} presents results on the spatio-temporal evolution of the 
exact solution for various values of $\kappa$. These numerical results were 
obtained by using a fourth-order accurate, central finite difference scheme 
for the (one-dimensional) Laplacian on a computational domain with half-width 
$L=50$ and resolution $\Delta x=0.1$. Note that Fig.~\ref{fig2}(d) demonstrates 
an unstable solution (i.e., a blowup case), and the time integration was stopped 
when the full-width-at-half-maximum (FWHM) was less than $\Delta x$ (for instance,
the blow-up time happens at a later time $t_{\mathrm{blowup}}\approx 474$ in this 
panel, as per the discretization employed herein.).

%
%
\section{\label{s:conclusions}Conclusions}

In this work, we have considered the stability of the trapped soliton solution 
of a generalized Manakov system of two coupled nonlinear \Schrodinger\ equations 
with the particular constraint $\psi_2(x,t) = \psi_1(-x,t)$ both numerically and 
analytically, and in a 2 and 4 collective coordinate (2CC and 4CC) approximation 
for arbitrary nonlinearity parameter $\kappa$. We were able to show that this system 
was equivalent to a nonlocal nonlinear \Schrodinger\ equation derivable from a nonlocal 
action. We found by a variety of methods, that the stability to width changes of this 
NNLSE had exactly the same behavior as the counterpart single component NLSE. That is, 
for $\kappa < 2$ there is stability, for $\kappa >2$ there is instability (either collapse 
or blowup), and at $\kappa=2$, there is a critical mass above which there is an instability. 
The exact solution is related to the critical mass. 
Unlike the NLSE which has Galilean invariance and a conserved momentum which can be nonzero, 
the NNLSE has zero value for the conserved momentum as well as having some potentially 
complex conserved normalization factors. This leads to a different response to shifting the 
initial conditions on the wave function to be slightly different than zero.  
When we add a nonzero `chirp' term to the phase of the exact solution, we find that this 
adds extra energy to the soliton which then results in the soliton width having oscillations 
which are in accordance with this extra energy. For the NNLSE parity is conserved and the total 
momentum is zero. The system does not have Galilean invariance. When we shift the soliton slightly 
from the origin, the ensuing oscillations are reasonably well captured by the 4CC approximation
which violates parity conservation. In conclusion, we have mapped out the stability regions for 
the NNLSE with arbitrary nonlinearity parameter $\kappa$ and have studied the response of the exact 
solutions both analytically (in the realm of a collective coordinate approximation) as well as 
numerically, and found the two approaches are in accordance. 

\ack
We thank M.~Lakshmanan (Bharathidasan University) for useful correspondence. 
EGC, FC, and JFD would like to thank the Santa Fe Institute and the Center 
for Nonlinear Studies at Los Alamos National Laboratory for their hospitality. 
AK is grateful to Indian National Science Academy (INSA) for awarding him 
INSA Senior Scientist position at Savitribai Phule Pune University, Pune, 
India. The work of AS was supported by the U.S.\ Department of Energy. 

%
%
\appendix
%
%
%
%
\section{\label{s:integrals}Useful integrals and identities}

\begin{figure}[t]
\centering
\includegraphics[width=0.95\columnwidth]{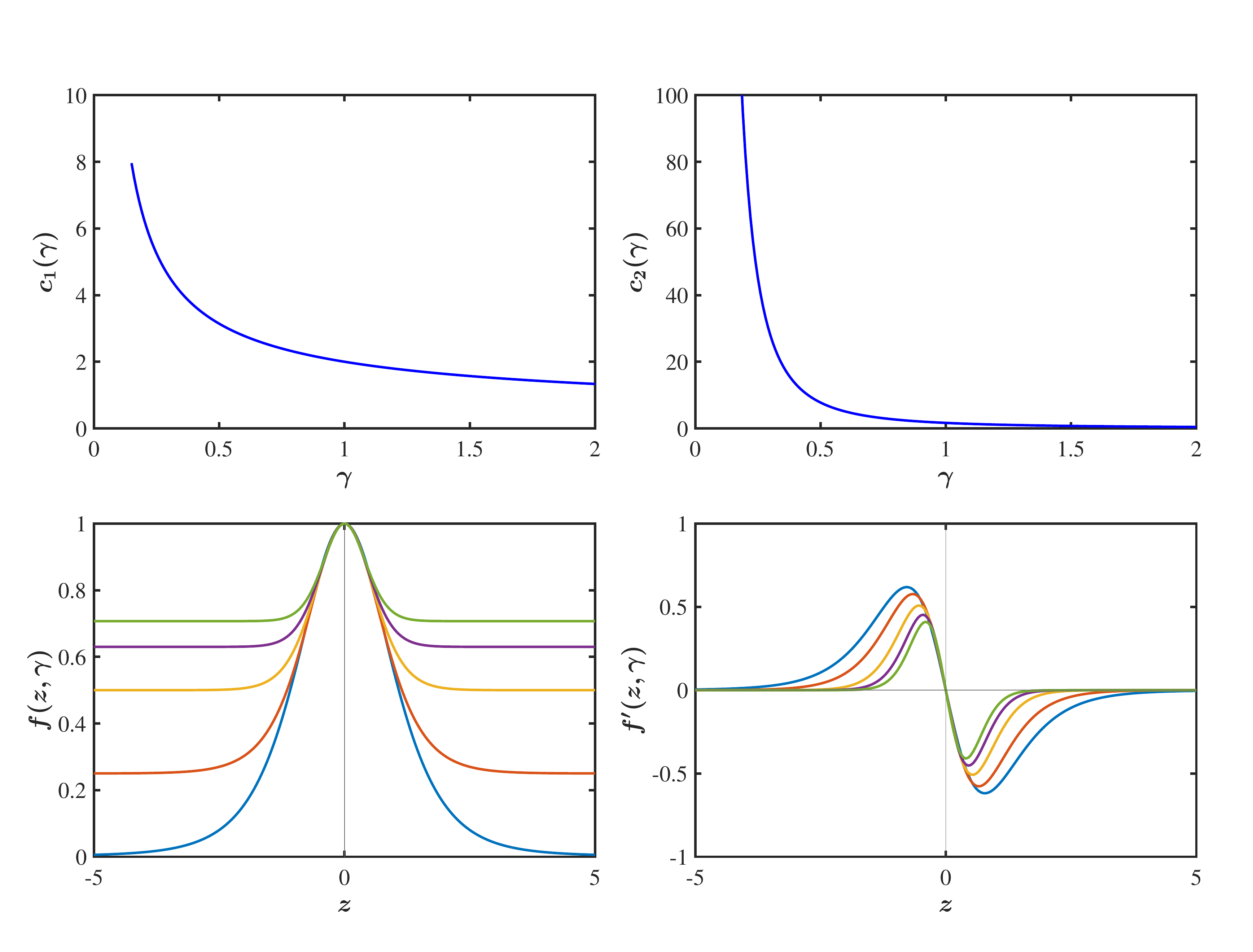}
\caption{\label{fig:fzgamma}
The top panels present $c_1(\gamma)$ (left panel) and $c_2(\gamma)$ 
(right panel) as a function of $\gamma$ (for their definition, see 
Eqs.~\eqref{c1def} and~\eqref{c2def}, respectively). The bottom panels 
demonstrate the dependence of $f(z,\gamma)$ (left panel), and $f'(z,\gamma)$ 
(right panel) on $z$, and for various values of $\gamma = 0.1, 0.5, 1, 1.5$, 
and $2$ depicted by solid blue, orange, yellow, purple, and green lines, 
respectively (see Eqs.~\eqref{fdef-II} and~\eqref{fprime}, respectively).
}
\end{figure}

\begin{subeqnarray}\label{cdefs}
   c_1(\gamma)
   &=
   \int_{-\infty}^{+\infty} \!\!\!\! \dd{z} \sech^{2\gamma}(z)
   =
   \frac{\sqrt{\pi} \, \Gamma[\gamma]}{\Gamma[\gamma+1/2]} \>,
   \label{c1def} \\
   c_2(\gamma)
   &=
   \int_{-\infty}^{+\infty} \!\!\!\! \dd{z} z^2 \, \sech^{2\gamma}(z)
   \label{c2def} \\
   &=
   2^{2\gamma - 1}\,
   {}_4F_3[\gamma,\gamma,\gamma,2\gamma;1+\gamma,1+\gamma,1+\gamma;-1] / \gamma^3 \>, 
   \notag \\
   c_3(\gamma)
   &=
   \int_{-\infty}^{+\infty} \!\!\!\! \dd{z} \sech^{2\gamma}(z) \,
      \tanh^2(z)
   \label{c3def} \\
   &=
   c_1(\gamma) - c_1(\gamma+1)
   =
   \frac{c_1(\gamma)}{2 \gamma + 1} \>.
   \notag
\end{subeqnarray}
A useful result is
\begin{equation}\label{Identity-1}
   \frac{c_1(\gamma+1)}{c_1(\gamma)}
   =
   \frac{2 \gamma}{2 \gamma + 1} \>.
\end{equation}
We have defined the integral $f(z,\kappa)$ in \eqref{fdef}:
\begin{equation}\label{fdef-II}
   f(z,\gamma)
   =
   \frac{2 \gamma +1}{2^{1/\gamma + 2} \, \gamma \, c_1(\gamma)} 
   \tint \dd{y} 
   [\, \sech^{2 \gamma}(y - z) +  \sech^{2 \gamma}(y + z) \,]^{1/\gamma+1} \>.
\end{equation}
Then
\begin{equation}\label{fzero}
   f(0,\gamma)
   =
   \frac{2 \gamma +1}{2 \gamma \, c_1(\gamma)}   
   \tint \dd{y} \sech^{2 \gamma + 2}(y)
   =
   \frac{2 \gamma +1}{2 \gamma} \, \frac{c_1(\gamma+1)}{c_1(\gamma)}
   = 1 \>.
   \notag
\end{equation}
Plots of $f(z,\gamma)$ are shown in the left panel in Fig.~\ref{fig:fzgamma}.
The derivative of $f(z,\gamma)$ wrt $z$ is given by
\begin{eqnarray}
   \fl
   f'(z,\gamma)
   &=
   \frac{(\gamma + 1)( 2 \gamma + 1)}{2^{1/\gamma + 1} \gamma \, c_1(\gamma) } 
   \tint \dd{y}
   [\, \sech^{2 \gamma}(y - z) +  \sech^{2 \gamma}(y + z) \,]^{1/\gamma} \>
   \label{fprime} \\
   \fl
   & \hspace{5em}
   \times
   [\, 
      \sech^{2 \gamma}(y - z) \tanh(y - z) 
      - 
      \sech^{2 \gamma}(y + z) \tanh(y + z) \,
   ] \>,
   \notag
\end{eqnarray}
where $f'(0,\gamma) = 0$. Plots of $f'(z,\gamma)$ are shown in the right panel 
in Fig.~\ref{fig:fzgamma}. A short calculation using Mathematica gives
\begin{equation}\label{fppzero}
   f''(0,\gamma)
   =
   - 4 \, \frac{\gamma+1}{2\gamma+3} \>.
\end{equation}
Expanding $f(z,\gamma)$ about the origin gives
\begin{subeqnarray}\label{fexpand}
   \fl
   f(z,\gamma)
   &=
   f(0,\gamma) + f'(0,\gamma) \, z + \frac{1}{2} \, f'(0,\gamma) \, z^2 + \cdots
   =
   1 - 2 \, \frac{\gamma+1}{2\gamma+3} \, z^2 + \cdots \>,
   \label{fexpand-a} \\
   \fl
   f'(z,\gamma)
   &=
   - 4 \, \frac{\gamma+1}{2\gamma+3} \, z + \cdots \>. 
   \label{fexpand-b}
\end{subeqnarray}
%
%
%
%
\section{\label{s:intexact}Moments and collective coordinates}

The variational wave function of Eq.~\ef{4CCAnsatz} is of the form
\begin{subeqnarray}\label{4CCAnsatz-II}
   \psi(x,t)
   &=
   \Bigl [\, \frac{M}{G(t) \, c_1(\gamma)} \, \Bigr ]^{1/2} \, 
   \sech^{\gamma}\Bigl [\,\frac{x - q(t)}{G(t)} \, \Bigr] \, 
   \rme^{\rmi \phi(x,t)} \>,
   \label{varpsi4CC-II} \\
   \phi(x,t)
   &=
   p(t) \,(x - q(t)) + \Lambda(t) \, (x - q(t))^2 \>.
   \label{varphi4CC-II}
\end{subeqnarray}
Let us define the $n^{\mathrm{th}}$ moment of the density distribution by
\begin{eqnarray}\label{Moments}
   M_n(t)
   &=
   \int_{-\infty}^{+\infty} \!\!\! \dd{x} x^n \, |\, \psi(x,t) \,|^2 
   \\
   &=
   \frac{M}{G(t) \, c_1(\gamma)}
   \int_{-\infty}^{+\infty} \!\!\! \dd{x} x^n \, \sech^{2 \gamma}[\, (\, x - q(t) \,)/G(t) \,] 
   \notag \\
   &=
   \frac{M}{c_1(\gamma)}
   \int_{-\infty}^{+\infty} \!\!\! \dd{y} \bigl [\, G(t) \, y + q(t) \, \bigr ]^n \,
   \sech^{2 \gamma}(y) \>.
   \notag
\end{eqnarray}
Explicitly, we find:
\begin{subeqnarray}\label{AlldenMoments}
   M_0(t) &= M \>,
   \label{moment-0} \\[5pt]
   M_1(t) &= M \, q(t) \>,
   \label{moment-1} \\
   M_2(t) &= M \,
   \Bigl [\, q^2(t) + G^2(t) \, \frac{c_2(\gamma)}{c_1(\gamma)} \, \Bigr ] \>,
   \label{moment-2}
\end{subeqnarray}
from which we can find $q(t)$ and $G(t)$.  The zeroth moment, i.e., the mass of
the soliton must be conserved. In a similar way, let us define the $n^{\mathrm{th}}$ 
moment of the momentum distribution by
\begin{eqnarray}\label{pmoment}
   \fl
   P_n(t)
   &=
   \frac{1}{2 \rmi} \int_{-\infty}^{+\infty} \!\! \dd{x} x^n \,
   \Bigl [ \,
      \psi^{\ast}(x,t) \, \pdv{\psi(x,t)}{x}
      -
      \pdv{\psi^{\ast}(x,t)}{x} \, \psi(x,t) \,
   \Bigr ]
   \\
   \fl
   &=
   \int_{-\infty}^{+\infty} \!\! \dd{x} x^n \,
   \Imag{ \psi^{\ast}(x,t) \, \pdv{\psi(x,t)}{x} }
   =
   \int_{-\infty}^{+\infty} \!\! \dd{x} x^n \, \pdv{\phi(x,t)}{x} \, |\, \psi(x,t) \,|^2
   \notag \\
   \fl
   &=
   \int_{-\infty}^{+\infty} \!\!\! \dd{x} x^n \,
   \{\, p(t) + 2 \, \Lambda(t) \, [x - q(t)] \,\} \, |\, \psi(x,t) \,|^2 
   \notag \\
   \fl
   &=
   \frac{M}{G(t) \, c_1(\gamma)} \int_{-\infty}^{+\infty} \!\!\! \dd{x} x^n \,
   \{\, p(t) + 2 \, \Lambda(t) \, [x - q(t)] \,\} \, \sech^{2\gamma}[(x - q(t))/G(t)]
   \notag \\
   \fl
   &=
   \frac{M}{c_1(\gamma)} \int_{-\infty}^{+\infty} \!\!\! \dd{y}
   [\, q(t) + G(t) \, y\,]^n \,
   [\, p(t) + 2 \, \Lambda(t) \, G(t) \, y \,] \, \sech^{2\gamma}(y)\>.
   \notag
\end{eqnarray}
Explicitly, we find:
\begin{subeqnarray}\label{AllPMoments}
   P_0(t)
   &=
   M \, p(t) \>,
   \label{P0-moment} \\
   P_1(t)
   &=
   M \, 
   \Bigl [\,
      p(t) \, q(t) 
      + 
      2 \, \Lambda(t) \, G^2(t) \, \frac{c_2(\gamma)}{c_1(\gamma)} \,
   \Bigr ] \>.
   \label{P1-moment}
\end{subeqnarray}
This way, $p(t)$ can be obtained from \eqref{P0-moment}, and $\Lambda(t)$ from:
\begin{equation}\label{getLambda}
   \Lambda(t)
   =
   \frac{1}{2 \, G^2(t)} \,
   \Bigl [ \, \frac{P_1(t)}{M} - p(t) \, q(t) \, \Bigr ] \, 
   \frac{c_1(\gamma)}{c_2(\gamma)} \>.
\end{equation}
%

%
%
\section*{References}
%

\bibliography{johns.bib}

\end{document}